\documentclass[%
 reprint,
superscriptaddress,
 showkeys,
 amsmath,amssymb,
 aps,
 pra,
]{revtex4-2}

\usepackage{graphicx}
\usepackage{dcolumn}
\usepackage{bm}
\usepackage{nicefrac}
\usepackage{hyperref}
\usepackage{threeparttable}
\usepackage{booktabs}
\usepackage{notoccite}
\usepackage{siunitx}
\usepackage{xr}
\usepackage{cleveref}
\usepackage{xr}
\usepackage{soul}

\renewcommand{\eqref}[1]{Eq.~(\ref{#1})}
\newcommand{\figref}[1]{Fig.~\ref{#1}}

\makeatletter
\newcommand*{\addFileDependency}[1]{
  \typeout{(#1)}
  \@addtofilelist{#1}
  \IfFileExists{#1}{}{\typeout{No file #1.}}
}
\makeatother
\newcommand*{\myexternaldocument}[1]{
    \externaldocument{#1}
    \addFileDependency{#1.tex}
    \addFileDependency{#1.aux}
}
\myexternaldocument{supplementary}

\newcommand{\rev}[1]{#1}

\begin{document} 

\title{Active learning of the thermodynamics--dynamics tradeoff in protein condensates} 

\author{Yaxin An}
\altaffiliation{Present address: Department of Chemical Engineering, Louisiana State University, Baton Rouge, LA 70803 USA}
\affiliation{%
  Department of Chemical and Biological Engineering, Princeton University, Princeton, NJ 08544 USA\\
}%
\affiliation{%
  Department of Chemistry, Princeton University, Princeton, NJ 08544 USA
}%
\author{Michael A. Webb}%
 \email{mawebb@princeton.edu}
\affiliation{%
  Department of Chemical and Biological Engineering, Princeton University, Princeton, NJ 08544 USA\\
}%
\author{William M. Jacobs}%
 \email{wjacobs@princeton.edu}
\affiliation{%
  Department of Chemistry, Princeton University, Princeton, NJ 08544 USA
}%

\date{\today}

\begin{abstract}
Phase-separated biomolecular condensates exhibit a wide range of dynamical properties, which depend on the sequences of the constituent proteins and RNAs. However, it is unclear to what extent condensate dynamics can be tuned without also changing the thermodynamic properties that govern phase separation. Using coarse-grained simulations of intrinsically disordered proteins, we show that the dynamics and thermodynamics of homopolymer condensates are strongly correlated, with increased condensate stability being coincident with low mobilities and high viscosities. We then apply an ``active learning'' strategy to identify heteropolymer sequences that break this correlation. This data-driven approach and accompanying analysis reveal how heterogeneous amino-acid compositions and non-uniform sequence patterning map to a range of independently tunable dynamical and thermodynamic properties of biomolecular condensates. Our results highlight key molecular determinants governing the physical properties of biomolecular condensates and establish design rules for the development of stimuli-responsive biomaterials. 
\end{abstract}

\maketitle 

\section*{Introduction}

Biomolecular condensates---also known as ``membrane-less organelles''---are self-organized structures within the cytoplasm and nucleoplasm of living cells~\cite{brangwynne2009germline,gomes2019molecular,mittag2022conceptual}.
Condensates play diverse roles in a wide variety of biological processes~\cite{kang2022llps,he2023phase,shin2017liquid}, in large part because of their ability to concentrate proteins, RNAs, and small molecules in chemically specific environments~\cite{deng2017microfluidic,kojima2018membraneless,shin2017liquid,he2023phase}.
Nonetheless, the ability of condensates to tune the rates of biochemical reactions is not only determined by their biochemical compositions.
Dynamical properties, such as viscosities and molecular mobilities, also affect how condensates exchange biomolecules with the rest of the cell~\cite{hazra2021biophysics,zhao2020phase,alshareedah2021programmable} and interact with other intracellular structures, such as chromatin~\cite{cai2023through} and cytoskeletal components~\cite{price2021intrinsically}.
Dynamical properties are also directly relevant to condensate assembly and disassembly kinetics~\cite{mediani2021hsp90} and irreversible aging processes~\cite{agarwal2021intrinsically}.
Recent experiments have demonstrated that dynamical properties vary widely across different condensates, both \textit{in vitro} and \textit{in vivo}~\cite{alshareedah2021programmable,wei2017phase}.

Many condensates are believed to form spontaneously as a result of phase separation taking place close to a local thermodynamic equilibrium~\cite{brangwynne2009germline,feric2016coexisting}.
The thermodynamic stability of a condensate, which can be quantified by the critical temperature of a biopolymer solution~\cite{dignon2018relation} or by the saturation concentration required for phase separation~\cite{hazra2021biophysics,das2018coarse}, is an equilibrium property related to the multiplicity and strengths of the ``multivalent'' interactions among the biomolecules comprising a condensate~\cite{hazra2021biophysics,das2020comparative,fung2018idps}.
Intuitively, one might expect that the thermodynamic stability of a condensate is correlated with its dynamical properties, which also emerge from the collective effects of biomolecular interactions in the condensed phase.
This anticipated relationship represents a ``tradeoff'' between stability and dynamics, since stronger thermodynamic driving forces are expected to correlate with lower molecular mobilities~\cite{R:1985_Harding_concentration}.
Experimental evidence suggests that such a correlation exists for certain ribonucleic condensates, such as Arg/Gly-rich repeat protein/RNA condensates~\cite{alshareedah2021programmable} and LAF-1/RNA condensates~\cite{wei2017phase}.
However, understanding the general relationship between thermodynamic stability and the internal dynamics of biomolecular condensates requires additional and systematic exploration~\cite{laghmach2022rna,wadsworth2022rnas}.

Determining whether a stability--dynamics tradeoff extends to phase-separating biopolymers is challenging due to the enormous diversity of biopolymers that can form condensates.
Here, we focus on condensates composed of intrinsically disordered proteins (IDPs), a class of proteins that tend not to adopt stable secondary and tertiary structures and are known to be essential components of many naturally occurring condensates~\cite{oldfield2014intrinsically}.
To study the properties of IDP sequences, we employ a coarse-grained (CG) model that represents IDPs at residue-level resolution~\cite{regy2021improved}.
Models of this type have been shown to reproduce experimental measurements of both single-molecule and condensed-phase properties~\cite{dignon2018relation,dignon2019simulation,dignon2018sequence,joseph2021physics}, suggesting that CG IDP models capture essential physics of disordered polypeptides and retain sufficient chemical specificity to predict differences among condensates composed of various IDP sequences.
Thus, CG IDP models seem well-suited to systematically explore the relationship between IDP stability and condensate dynamics. 

In this article, we combine molecular dynamics simulations and machine learning to navigate the sequence space~\cite{R:2020_Webb_Targeted,R:2021_Gormley_Machine} of IDPs and elucidate the relationship between the phase behavior and internal dynamics of single-component IDP condensates.
We first establish a strong correlation between proxies for condensate stability and dynamics in homomeric polypeptides, which is understood using simple physical arguments.
We then deploy a computationally efficient machine-learning strategy based on ``active learning''\cite{R:2018_Smith_Less,R:2019_Ramprasad_Active_MRSCommunications,shahriari2015taking} to identify IDP sequences that break this correlation, exhibiting faster internal dynamics than homomeric polypeptides that assemble into equally stable condensates.
In this way, we identify ``Pareto-optimal'' IDP sequences, meaning that sequence perturbations cannot enhance the dynamics further without reducing the stability of the condensate.
Finally, we examine sequence features of Pareto-optimal sequences and perform a counterfactual analysis\cite{R:2022_Wellawatte_Model,R:2023_Wellawatte_Perspective} to identify the sequence determinants of the limiting thermodynamics--dynamics tradeoff.
Taken together, our results demonstrate how sequence design can be used to tune thermodynamic and dynamical properties independently in the context of biomolecular condensates.

\section*{Results}

\subsection*{Thermodynamic and dynamic properties of homomeric polypeptides are strongly correlated}
\label{sec:homomeric}

To establish a baseline expectation for the thermodynamics--dynamics tradeoff in IDP condensates, we begin by studying a collection of homopolymeric sequences at \SI{300}{K}.
\rev{Specifically, we consider homopolymers consisting of each amino-acid type with chain lengths of $N=20$, $30$, $40$, or $50$ monomers.}
Molecular dynamics simulations are performed using the CG model developed by Mittal and coworkers~\cite{regy2021improved,dignon2018sequence}, which treats IDPs at the single-residue level in implicit \rev{aqueous} solvent.
Amino-acid residues interact via a combination of bonded and nonbonded pair potentials, the latter of which account for both electrostatic interactions and short-ranged hydrophobic forces.
\rev{We use a Debye screening length of \SI{1}{nm}, corresponding to an ionic strength of \SI{0.1}{M}.
See ``Model of intrinsically disordered proteins'' in Materials and Methods for further details.}

\begin{figure}[!htbp]
  \centering \includegraphics[width=\columnwidth]{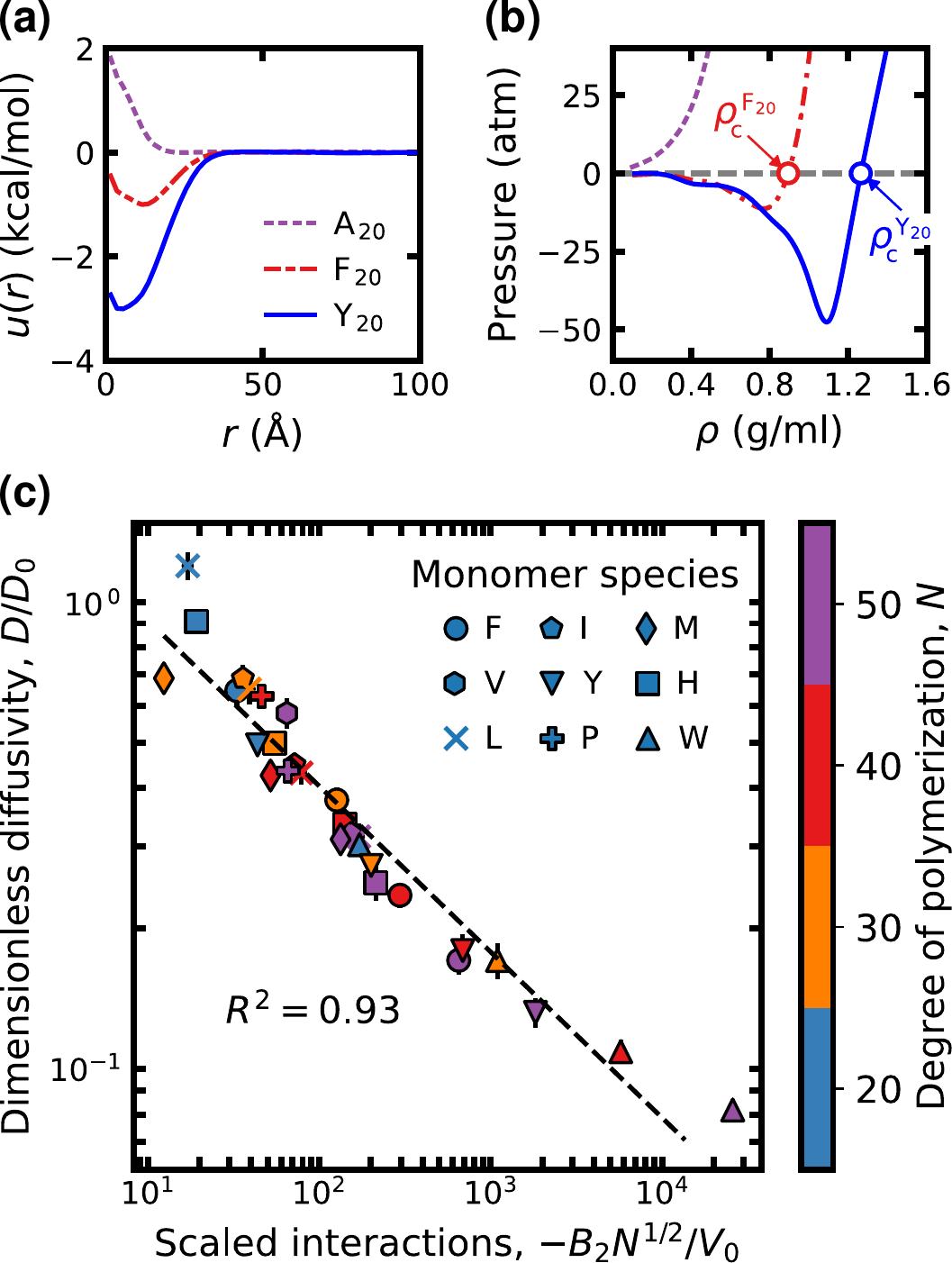}
  \caption{\textbf{Coarse-grained simulations of homomeric polypeptides predict a strong relationship between condensate stability and internal dynamics.}
    \textbf{(a)}~The potential of mean force, $u(r)$, between the centers of mass of two polymer chains at \SI{300}{K}.  Three example curves are shown for net repulsive, weakly attractive, and strongly attractive 20-mers, $\text{A}_{20}$, $\text{F}_{20}$, and $\text{Y}_{20}$, respectively.
    \textbf{(b)}~Empirical equation-of-state curves at \SI{300}{K} for the three homomeric polypeptides in panel \textbf{a}.  Only two of these 20-mers, $\text{F}_{20}$ and $\text{Y}_{20}$, are predicted to phase separate to form condensed phases with the indicated densities $\rho_{\text{c}}^{\text{F}_{20}}$ and $\rho_{\text{c}}^{\text{Y}_{20}}\!$, respectively.
    \textbf{(c)}~The dimensionless self-diffusion coefficients, $D/D_0$, of homopolymers in the condensed phase are anticorrelated with their scaled dimensionless second-virial coefficients, $-B_2 N^{1/2}/ V_0$.  Points are shown on a log--log plot for the 29 homopolymers (out of 80 simulated) that phase separate at \SI{300}{K}.  Self-diffusion coefficients are obtained from canonical-ensemble simulations conducted at the condensed-phase density, $\rho_{\text{c}}$, determined from the equation-of-state analysis.  Statistical errors reflect the standard error of the mean and are comparable to the symbol size.
} 
  \label{fig:homopolymer}
\end{figure}

Throughout this work, we utilize the second-virial coefficient, $B_2$, as a proxy for the thermodynamic stability of the condensed phase \rev{for IDPs that undergo phase separation.}
$B_2$ quantifies the net attractive or net repulsive intermolecular interactions of a pair of molecules in dilute solution~\cite{hazra2022affinity}.
\rev{It is well established that $B_2$ can also be used to predict the critical point of simple atomic and molecular fluids~\cite{mcquarrie1976statistical}, colloidal suspensions~\cite{vliegenthart2000predicting,tuinier2000depletion}, and homopolymer solutions~\cite{rubinstein2003polymer}.
A strong correlation between $B_2$ and the critical temperature for IDP phase separation has also been reported, albeit for a limited set of IDP sequences~\cite{dignon2018relation}.
In all these systems, a negative $B_2$ value---implying net attractive interactions---is necessary for phase separation to occur.
Although this condition turns out to be insufficient to guarantee phase separation in heteropolymer solutions~\cite{R:2023_Rekhi_Role,R:2023_Panagiotopoulos_Phase}, we find that $B_2$ nonetheless correlates strongly with the difference between the coexisting phase densities for IDP sequences that do phase separate, as we demonstrate using molecular dynamics simulations below.}
In practice, we compute $B_2$ by calculating the potential of mean force, $u(r)$, between the centers of mass of two polymer chains at \SI{300}{K} using adaptive biasing force simulations~\cite{comer2015adaptive} (\figref{fig:homopolymer}a; see also ``Physical property calculations'' in Materials and Methods).
$B_2$ is then obtained from the equation
\begin{equation}
  \label{eq:B2}
  B_2 = 2\pi \int_0^\infty dr\,r^2 \left[1 - e^{-\beta u(r)}\right],
\end{equation}
where $r$ represents the distance between the center of mass of each chain and $\beta \equiv 1/k_{\text{B}}T$.
For convenience, we report $B_2$ relative to a reference volume, $V_0=\SI{5529}{\angstrom^3}$, equal to the pervaded volume of an ideal polymer chain, ${V_0 = (4\pi/3)b_0^3 (N/6)^{3/2}}$~\cite{rubinstein2003polymer}, with the CG equilibrium bond length, $b_0=$ \SI{3.8}{\angstrom}, and a chain length of $N=50$.

\rev{Next, we develop a method to confirm whether an IDP sequence phase separates and, if so, to determine dynamical properties in the condensed phase.}
We perform an equation-of-state (EOS) analysis to approximate the condensed-phase density, $\rho_{\text{c}}$, for polymers with negative second-virial coefficients (\figref{fig:homopolymer}b; see also ``Physical property calculations'' in Materials and Methods).
\rev{The EOS is a relationship between the osmotic pressure and volume of the polymer solution at constant temperature.
In macroscopic systems, the pressure is a monotonically increasing function of the density at constant temperature.
However, in finite-size simulations, the EOS exhibits a non-monotonic ``van der Waals loop''~\cite{mcquarrie1976statistical,binder2012beyond} at temperatures for which phase separation occurs; fundamentally, this non-monotonicity arises due to the inability of a small system to form a stable interface between coexisting bulk phases.
We therefore use short simulations of 100 polymer chains at \SI{300}{K} to compute the pressure as a function of density, and we identify non-monotonic behavior as evidence of a phase-separated region.}
Making the approximation that the coexistence pressure is near zero, corresponding to a dilute low-density phase, we approximate the condensed-phase density, $\rho_{\text{c}}$, as the highest-density root of a non-monotonic EOS.
By contrast, we consider a strictly non-negative EOS as evidence that phase separation does not occur at \SI{300}{K}.
\rev{Following this approach, we find that 29 out of the 80 homomeric polypeptides phase separate at 300 K.}
We later show that this method, which is amenable to high-throughput simulations, is accurate to within a few percent of $\rho_{\text{c}}$ values determined via direct-coexistence slab simulations~\cite{dignon2018sequence}.

Finally, we use the self-diffusion coefficient, $D$, of a tagged chain within the bulk condensed phase as a simple measure of the internal condensate dynamics.
We calculate $D$ from the mean-squared displacement of a tagged chain in a canonical-ensemble simulation conducted at the condensed-phase density, $\rho_{\text{c}}$ (see ``Physical property calculations'' in Materials and Methods).
This quantity is negatively correlated with the viscosity of the condensed phase in phase-separated homopolymer solutions (Fig.~S1 in the Supplementary Materials).
However, the greater statistical uncertainties in the viscosity calculations make $D$ the more practical choice for quantifying internal condensate dynamics in high-throughput simulations.
We report self-diffusion coefficients relative to a reference value, $D_0 = \SI{0.42E-9}{m^2/s}$, which corresponds to an ideal Rouse chain with chain length $N=50$, $D_0 = \tau k_{\text{B}}T / N \bar{M}$~\cite{rubinstein2003polymer}, where $\tau = \SI{1}{ps}$ is the damping time in our Langevin simulations and $\bar M = \SI{118}{g/mol}$ is the average molecular weight of the 20 amino acid types.

Plotting $D/D_0$ versus $B_2/V_0$, we observe a strong negative correlation between these quantities across all homopolymeric sequences that are determined to phase separate (\figref{fig:homopolymer}c).
Surprisingly, this correlation is stronger than the direct correlation between the condensed-phase density, $\rho_{\text{c}}$, and either of these quantities alone (Fig.~S2 in the Supplementary Materials).
This indicates that even though the condensed-phase density at \SI{300}{K} cannot be predicted solely on the basis of the second-virial coefficient, this quantity nonetheless captures the key physics necessary to predict the internal dynamics of a homopolymer condensate \textit{at its equilibrium density}.

To rationalize the empirical relationship between $D/D_0$ and $B_2/V_0$, we consider a Rouse model~\cite{rubinstein2003polymer} of an unentangled polymer melt.
This model can describe the condensed phases in these simulations, which have densities on the order of \SI{1}{g/ml}.
In this model, the self-diffusion coefficient is inversely proportional to the total frictional force experienced by the chain, which typically follows an Arrhenius law for a thermally activated process~\cite{wadsworth2022rnas,rubinstein2003polymer}.
It is thus reasonable to relate the logarithm of the total friction to the reversible work required to separate a pair of chains, which is proportional to $\log(-B_2)$ in the limit of strong attractive interactions.
Accounting for the number of interacting chains within the pervaded volume of a tagged polymer~\cite{rubinstein2003polymer}, we therefore propose that the condensed-phase self-diffusion coefficient for phase-separating homopolymers should scale as $D \sim -N^{1/2} B_2$.
Empirically, we find that including the prefactor $N^{1/2}$ in the scaling relationship indeed improves the correlation across the 29 phase-separating homopolymers that we simulated (\figref{fig:homopolymer}c and Fig.~S3 of the Supplementary Materials).
We thus conclude that this simple model captures the essential relationship between the second-virial coefficients and condensed-phase self-diffusion coefficients of homomeric polypeptide sequences.

\subsection*{Short polypeptides from known intrinsically disordered regions do not form condensed phases}
\label{sec:short-polypeptides}

Having established a correlation between $D/D_0$ and $B_2/V_0$ for homomeric polypeptides, we next examine the behavior of heteromeric polypeptides based on extant IDP sequences.
To do so, we perform CG simulations of 1,266 short ($20 \le N \le 50$) polypeptide sequences from DisProt~\cite{R:2019_Hatos_DisProt}, a manually curated database of disordered regions of proteins.
\rev{This range of sequence lengths is used throughout this study to limit the total computational expense of these simulations.}
Although these polypeptides feature diverse sequence characteristics (Fig.~S4 in the Supplementary Materials), none of these polypeptide sequences form condensed phases according to the criteria set forth in the prior section.
\rev{This observation could be attributed to potential bias within the DisProt database towards soluble sequences~\cite{R:2016_Yu_Natural}, the imposition of an upper limit on sequence lengths probed in this study, which excludes many longer extant IDP sequences that are known to play important roles in intracellular phase separation~\cite{R:2019_Hatos_DisProt,li2020llpsdb}, and our focus on single-component systems (e.g., exclusion of RNAs that may be required to observe phase-separation~\cite{shin2017liquid}).}

Of all analyzed DisProt sequences with $20 \le N \le 50$, we find that the vast majority (c.a.~80\%) have positive $B_2$ (Fig.~S5 in the Supplementary Materials), while the remaining 255 sequences exhibit monotonically increasing EOS curves, indicating that phase separation does not occur at \SI{300}{K} when simulated using this CG IDP model.
Nonetheless, we note that some of these $B_2$ values are more negative than some of the phase-separating homomeric polypeptides considered in \figref{fig:homopolymer}.
\rev{This observation highlights the importance of performing the EOS analysis to confirm whether a polypeptide undergoes phase separation at \SI{300}{K}.}

\begin{figure*}[!htbp]
  \includegraphics[width=0.9\textwidth]{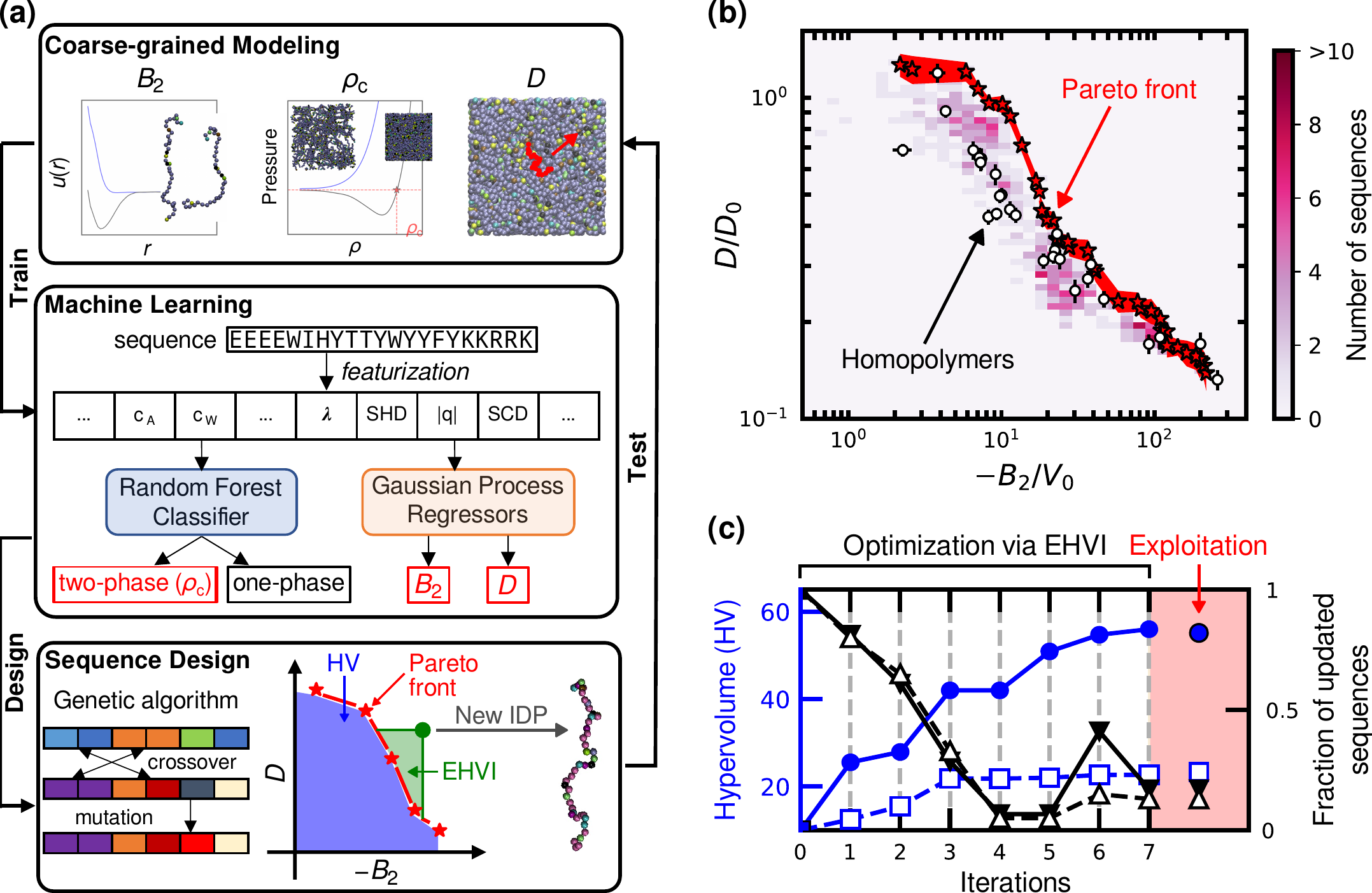}
  \caption {\textbf{Active-learning approach to heteromeric sequence design.}
    \textbf{(a)}~The workflow for designing new IDP sequences by integrating coarse-grained modeling, machine learning (ML), and genetic-algorithm-based optimization.  Active learning is an iterative process consisting of (i) performing molecular dynamics simulations to compute three physical properties of IDPs ($B_2$, phase behavior, and $D$); (ii) learning the predictive features of IDP sequences by training ML models to estimate these three physical properties; (iii) designing new sequences by maximizing a fitness function; and (iv) returning to step (i) to test the physical properties of the proposed IDP sequences.  The data set and ML models are subsequently updated via successive iterations.
    \textbf{(b)}~The relationship between the scaled condensed-phase self-diffusion coefficients, $D/D_0$, and scaled second-virial coefficients, $-B_2/V_0$, of all the designed IDP sequences that undergo phase separation.  The color of each grid element indicates the number of distinct sequences within that region.  The red highlighted area represents a one-standard-deviation region spanned by the Pareto-optimal heteromeric sequences (red stars).  Homopolymer sequences (white circles) are shown for comparison. 
    \textbf{(c)}~Convergence analysis of the active-learning approach.  The hypervolume under the Pareto front (blue symbols, left axis) increases with active-learning iterations, while the fraction of newly updated data points at the Pareto front (black symbols, right axis) tends to zero as the Pareto front converges. Filled symbols/solid lines show the convergence of the entire Pareto front, while open symbols/dashed lines describe the convergence of the Pareto front in the regime where $-B_2/V_0 \lesssim 50$. A final ``exploitation'' round (shaded region) of sequence design confirms convergence of the hypervolume, introducing only a few new sequences into the Pareto front representation.}
  \label{fig:active}
\end{figure*}

\subsection*{Active learning identifies limits of tunability for condensate thermodynamic-dynamic properties}
\label{sec:active-learning}

Since the lack of phase separation precludes analysis of condensed-phase dynamical properties using short DisProt sequences, we set out to generate novel heteromeric sequences and explore the thermodynamics--dynamics tradeoff beyond homopolymers.
To this end, we adopt an active learning scheme~\cite{R:2018_Smith_Less} that iteratively selects and simulates new polypeptides to probe this relationship methodically and efficiently (\figref{fig:active}a).
The selection of new polypeptides is guided by Bayesian optimization, which has been usefully deployed in materials design, including the design of heteropolymer sequences, by informing the next ``best'' sequence(s) to characterize~\cite{R:2020_Webb_Targeted,R:2020_Ferguson_Discovery_JCPB,R:2022_Tamasi_Machine,R:2022_Kosuri_Machine,R:2021_Jablonka_Bias,tran2020multi}.
Within the context of the
thermodynamics--dynamics tradeoff, the ``best'' sequences are those considered as \textit{Pareto optimal}~\cite{emmerich2008computation,R:2016_Zuluaga_pare}, meaning that no other polypeptide sequence can be found that simultaneously exhibits an increased $D$ and a decreased $B_2$. 

Our active learning scheme integrates Bayesian optimization with supervised machine-learning models and a genetic algorithm for designing polypeptides.
Polypeptides are chosen for characterization based on a policy of expected hypervolume improvement (EHVI)~\cite{R:2019_Yang_Multi}, which has previously been demonstrated to converge towards a true set of Pareto-optimal sequences~\cite{macleod2022self}.
These Pareto-optimal sequences are referred to as the \textit{Pareto front}.  
Optimizing EHVI over sequence space is facilitated by a genetic algorithm that leverages machine-learning (ML) models, trained using the results of prior CG simulations, to predict physical properties (i.e., phase behavior, $B_2$, and $D$) from sequence features of prospective polypeptides. 
In particular, the ML models employ a 30-dimensional \textit{feature vector} that incorporates the amino-acid composition~\cite{R:2022_Patel_Featurization,R:2023_Patel_Data} and ten additional sequence-level descriptors, including the sequence length, $N$; the average hydrophobicity per residue, $\bar\lambda$; the sequence hydropathy decoration parameter, SHD~\cite{zheng2020hydropathy}; the fraction of positively and negatively charged residues, $\bar{q}_+$ and $\bar{q}_-$; the net charge per residue, $|q|$; and the sequence charge decoration, SCD; the mean-field second virial coefficient~\cite{zheng2020hydropathy}, $B_2^{\text{MF}}$; the Shannon entropy, $S$; and the average molecular weight, $\bar M$.
All the ten sequence-level feature values are normalized and standardized as described in ``Design of novel intrinsically disordered proteins'' in Materials and Methods.
Each iteration of active learning involves training ML models with the current simulation data, generating 96 new polypeptide sequences via the genetic algorithm, and subsequently simulating these sequences according to the protocols introduced previously.
Additional technical details are reported in Materials and Methods, and complete descriptions of ML model generation and validation are provided in the Supplementary Materials (see Fig.~S6).

We find that this active-learning approach rapidly identifies a representative set of Pareto-optimal polypeptides (\figref{fig:active}b and Fig.~S7). 
First, using a ML model for $B_2$ trained using DisProt sequence data, we perform an initial round of simulations involving IDPs chosen to minimize $B_2$ only (Iteration 0).
These simulations establish relevant data for training a ML model for $D$, enabling subsequent multiobjective optimization via EHVI (see ``Design of novel intrinsically disordered proteins'' in Materials and Methods).
We quantitatively assess convergence of the active-learning approach by monitoring both the hypervolume (HV) under the current approximation of the Pareto front and the fraction of newly added Pareto-optimal sequences relative to the current number of Pareto-optimal points (\figref{fig:active}c).
Active learning initially leads to a rapid expansion of the Pareto front (Iterations 1--3), whereas subsequent iterations (Iterations 4--7) result in substantially diminished expansion and mostly inconsequential additions to the Pareto front, especially along the weakly-interaction portion ($-B_2/V_0 \lesssim 50$) of the front (dashed lines in \figref{fig:active}c).
In these latter rounds, the expansion of the front occurs primarily via the addition of sequences with extremely negative $B_2$ and low $D$.
We therefore halt optimization based on EHVI at this point and perform one last iteration of sequence design based on pure exploitation of the ML models, which confirms convergence of the Pareto front (\figref{fig:active}c).
The final data set includes 2,114 polypeptide sequences, including 35 Pareto-optimal sequences (denoted P1--P35, see Table~S1) that provide a good representation of the true Pareto front of heteromeric polypeptides.

By contrast with homomeric polypeptides, active learning identifies ``designed'' heteropolymer sequences spanning a range of $D$ values at fixed $B_2$ and vice versa (\figref{fig:active}b).
Although heteromeric polypeptides follow the same trend of decreasing $D$ with decreasing $B_2$ as found for as homomeric polypeptides, the data cannot be neatly collapsed by any obvious scaling.
For strongly attracting polypeptides ($-B_2/V_0 \gtrsim 50)$, all sequences possess limited condensed-phase mobility.
Most of these, including the homomeric sequences Y$_{40}$, W$_{40}$, and W$_{50}$, are within the statistical error region spanned by Pareto-optimal heteromeric sequences (\figref{fig:active}b).
However, more weakly interacting polypeptides at the Pareto front ($-B_2/V_0 \lesssim 50$) exhibit $D$ values that are roughly twice those of homopolymer sequences with the same $B_2$.
For example, at a fixed $B_2/V_0\approx10$ in ~\figref{fig:active}b, the reduced diffusion coefficient $D/D_0$ increases from $\sim 0.4$ to $\sim 0.9$.
(We note that, because the Bayesian optimization only targets expansion of the Pareto front, it is likely that the maximum dynamic range of $D$ values at fixed $B_2$ can exceed a factor of two.)
Thus, \figref{fig:active}b illustrates the overall prospects and limitations of tuning the thermodynamic--dynamic tradeoff via sequence design.
Specifically, when IDPs interact relatively weakly, sequence design might be used to fine-tune the dynamical properties of single-component IDP condensates that possess similar thermodynamic stabilities \rev{(here assessed by $-B_2$) or, alternatively, to design IDP condensates that likely} undergo phase separation below disparate critical temperatures while exhibiting similar dynamical properties.

\begin{figure*}[!htbp]
  \centering
  \includegraphics[width=0.9\textwidth]{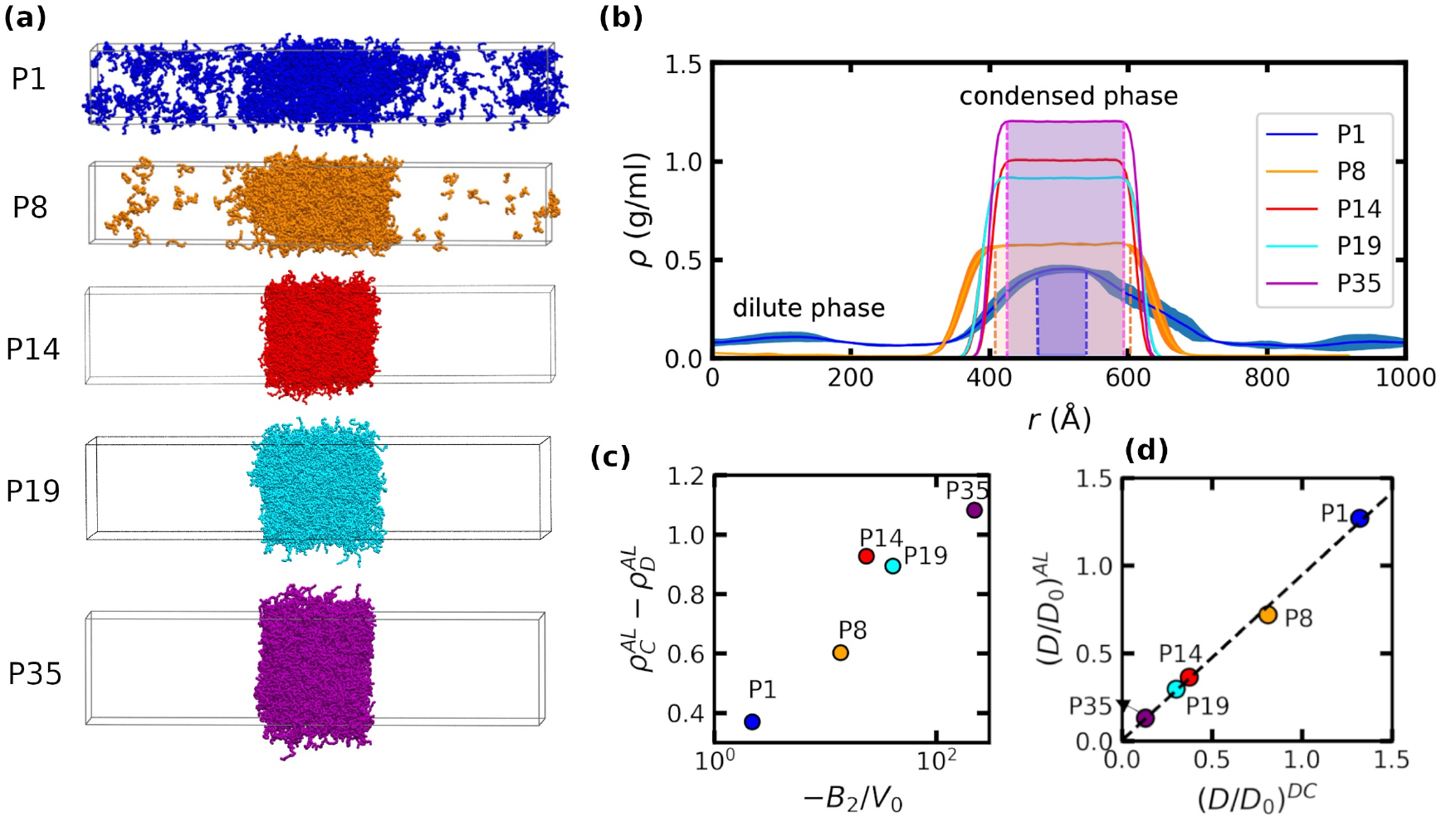}
  \caption{\textbf{Validation of Pareto-optimal IDP sequences using direct-coexistence simulations.}
    \textbf{(a)}~Snapshots of slab-geometry direct-coexistence simulations at \SI{300}{K} for five representative IDP sequences from the thermodynamics--dynamics Pareto front (see~\figref{fig:active}b): P1 (blue), P8 (orange), P14 (red), P19 (cyan), and P35 (purple).  Slab simulations of 1000 chains are performed in the canonical ensemble for \SI{4.5}{\us} (P1) and \SI{2}{\us} (P8, P14, P19, and P35).
    \textbf{(b)}~Average density profiles, measured orthogonally to the interface between the dilute and condensed phases, obtained from the simulations in panel \textbf{a}.  The final \SI{3}{\us} of the P1 trajectory and the final \SI{0.5}{\us} of the P8, P14, P19, and P35 trajectories are analyzed to compute these density profiles.  Shaded areas indicate each bulk condensed phase (except for P14 and P19 for clarity), defined as the region where variations in the average density are less than \SI{0.02}{g/ml}. 
    \rev{\textbf{(c)}~The correlation of the dimensionless second virial coefficient $-B_2/V_0$ and the density difference between the condensed and dilute phases, $\rho_C-\rho_D$.} \textbf{(d)}~Comparison of the dimensionless condensed-phase self-diffusion coefficients, $D/D_0$, computed via the equation-of-state method within the active-learning framework (AL) or via the direct-coexistence (DC) simulations shown in panel \textbf{a}. The dashed line shows a linear fit to the data with a coefficient of determination of $\text{R}^2 = 0.996$ and a slope of $0.94$.}
  \label{fig:validation}
\end{figure*}

\subsection*{Coexistence simulations validate Pareto-optimal sequences discovered via active learning}
\label{sec:direct-coexistence}

We validate the IDP sequences designed via active learning by performing large-scale molecular dynamics simulations of phase coexistence.
We carry out direct coexistence simulations of five representative sequences from the thermodynamics--dynamics Pareto front (P1, P8, P14, P19, and P35) using 1000 chains in a slab geometry~\cite{dignon2018relation, dignon2019simulation} at \SI{300}{K}.
Simulation durations of at least \SI{2}{\us} are required to reach equilibrium, resulting in the formation of an interface between equilibrium condensed and dilute phases (\figref{fig:validation}a). The interfaces between the coexisting dilute phases also become sharper as we move along the Pareto front from P1 to P35 (\figref{fig:validation}b), \rev{which is consistent with a trend towards increasing thermodynamic stability. 
\rev{Importantly, we also find that the differences between the condensed and dilute-phase densities are anticorrelated with $B_2$, which decreases from P1 to P35 (\figref{fig:validation}c).}
Both observations are consistent with the previously reported correlation between critical temperatures and second-virial coefficients in single-component IDP solutions~\cite{dignon2018relation} and support our use of $B_2$ as a proxy for the condensed-phase thermodynamic stability.}

Importantly, these simulations recapitulate the condensed-phase dynamical properties of Pareto-optimal sequences predicted by our active-learning approach.
From equilibrated simulations of phase coexistence, we determine the mean condensed-phase density and compute the self-diffusion coefficient of chains within the bulk of the condensed phase (i.e, away from the interface with the dilute phase).
The physical properties determined in this way compare favorably with the values determined from the EOS analysis (\figref{fig:validation}c and Fig.~S8 in the Supplementary Materials) and single-phase simulations of condensed-phase dynamical properties (\figref{fig:validation}d).
In all cases, quantities computed via these different methods differ by less than 10\% and typically agree within the statistical error.
This agreement demonstrates that the high-throughput simulations utilized by our active-learning approach are sufficiently accurate compared to more computationally expensive direct-coexistence simulations.
Furthermore, these results suggest that our active-learning protocol (\figref{fig:active}c) indeed converges to a representative sequences on the thermodynamics--dynamics Pareto front for this CG IDP model.

\subsection*{Sequence features change drastically along the thermodynamics--dynamics Pareto front}
\label{sec:pareto-features}

\begin{figure*}[!htbp]
  \includegraphics[width=\textwidth]{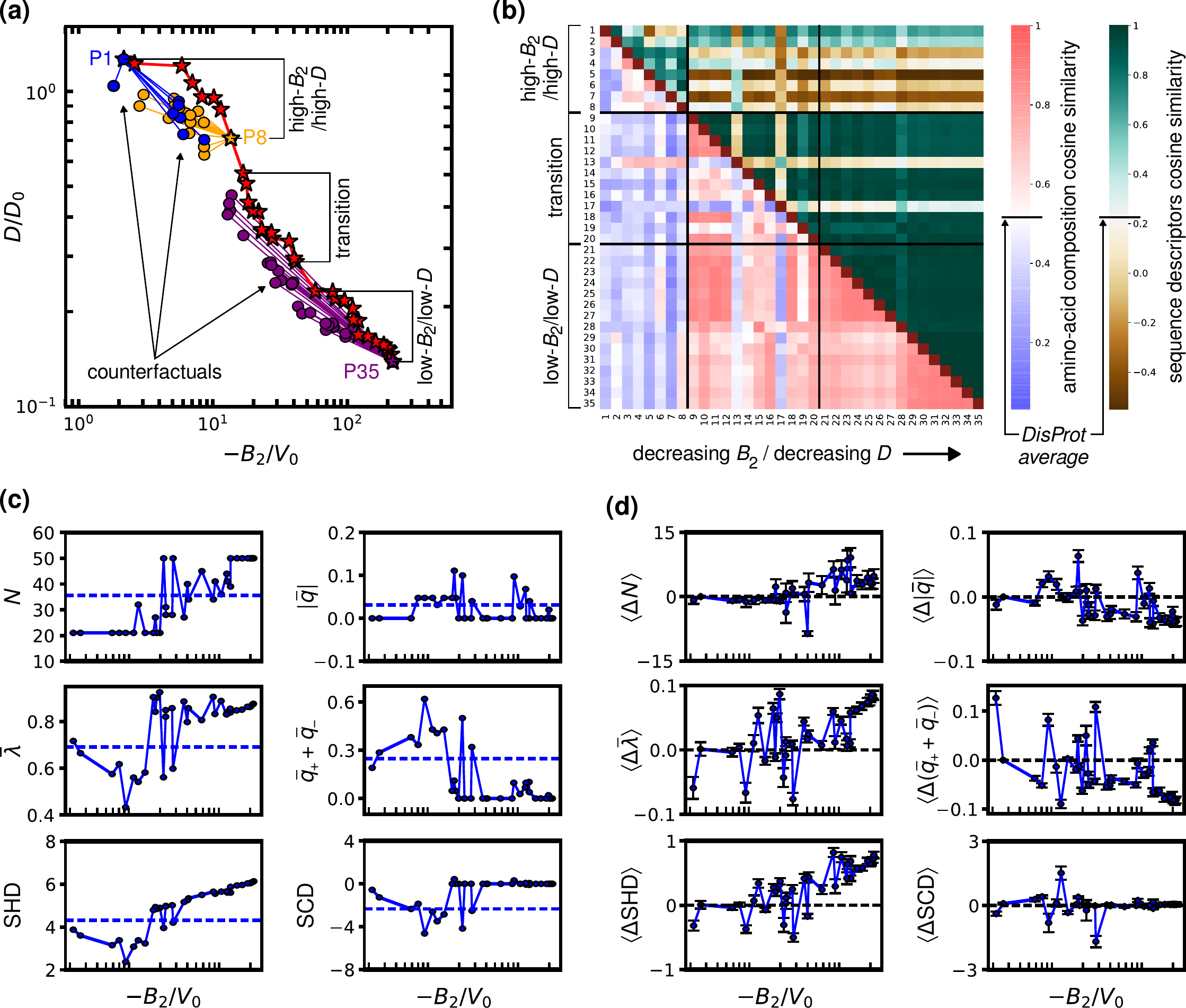}
  \caption{\textbf{Physicochemical features of sequences at the thermodynamics--dynamics Pareto front.}
    \textbf{(a)}~Representative IDP sequences along the Pareto front (stars) divide into three regimes.  Also shown are counterfactual sequences (circles) of three representative Pareto-optimal sequences, P1 (blue), P8 (orange), and P35 (purple); see text for the criteria used to select counterfactual sequences.
    \textbf{(b)}~Cosine similarities between the amino-acid composition component of the feature vectors (lower triangle, left color bar) and between the sequence-descriptor component of the feature vectors (upper triangle, right color bar) of the 35 representative Pareto-optimal sequences.
    \textbf{(c)} Variations of selected sequence descriptors (degree of polymerization, $N$; average hydrophobicity, $\bar\lambda$; hydropathy distribution, SHD; absolute net charge per residue, $|\bar{q}|$; the total charge, $\bar{q}_++\bar{q}_-$; and charge distribution, SCD) along the Pareto front.  Dashed lines indicate the mean value of each feature across all phase-separating sequences studied via active learning.
    \textbf{(d)} Average feature differences between each Pareto-optimal sequence and its counterfactuals, \eqref{eq:counterfactual}.  Error bars represent standard errors of the mean, and dashed lines indicate zero feature difference.}
  \label{fig:features}
\end{figure*}

To better understand what governs the thermodynamics--dynamics tradeoff  for heteromeric polypeptides (\figref{fig:active}b), we analyze the sequence characteristics of IDPs on the Pareto front (\figref{fig:features}a, stars).
An aggregate comparison among the Pareto-optimal polypeptides (P1--P35) is based on the pairwise cosine similarity of their feature vectors, where higher scores indicate greater similarity.
For clarity, we decompose sequence similarity into the contributions from \emph{(i)} the amino-acid composition (\figref{fig:features}b, lower triangle) and \emph{(ii)} the various sequence-derived descriptors (\figref{fig:features}b, upper triangle).
The average cosine similarities among pairs of Pareto-optimal sequences are \emph{(i)} 0.60 and \emph{(ii)} 0.56, respectively.
For reference, the average cosine similarities among pairs of DisProt sequences ($20 \leq N \leq 50$) are \emph{(i)} 0.57 and \emph{(ii)} 0.22; these reference values are indicated by the vanishing points of the color bars in \figref{fig:features}b.

Based on relative similarities, we propose that the Pareto-optimal sequences roughly divide into three groups: P1--P8, P9--P20, and P21--P35, outlining a high-$B_2$/high-$D$ regime, a transition regime, and a low-$B_2$/low-$D$ regime, respectively (top-left, middle, and bottom-right in \figref{fig:features}a,b).
\rev{While P1--P8 generally exhibit stronger composition and sequence-descriptor similarity than randomly selected pairs of disordered regions from DisProt, they significantly differ from the other groups.
Collectively, P9 to P35 tend to show strong composition and sequence-descriptor similarities with one another. 
However, the transition group (P9 to P20) has lower intra-group similarity compared to the low-$B_2$/low-$D$ group (P21 to P35) (e.g., see also dashed lines in Fig.~S9 of the Supplementary Materials), which motivates their distinction.}

\rev{However, we find that neither composition nor descriptor-based similarity alone guarantees similar macroscopic properties among Pareto-optimal sequences.
On the one hand, Pareto-optimal sequences can exhibit high variability in sequence characteristics despite occasionally possessing similar physical properties.
This is particularly true in the transition regime, which is typified by high intra-group dissimilarity; for instance, P13 and P17 differ significantly in terms of sequence descriptors relative to sequences that are directly adjacent on the Pareto front.}
This variability in sequence characteristics also suggests a higher degeneracy of near-Pareto-optimal polypeptides in the transition regime compared to the high-$B_2$/high-$D$ and low-$B_2$/low-$D$ regimes, where the characteristics of Pareto-optimal sequences tend to be more consistent.
On the other hand, sequences with high overall composition and sequence-descriptor similarity can exhibit vastly different macroscopic properties. 
For example, extremal sequences P1 and P35 have greater sequence-descriptor similarity than P1 does with several other sequences in the high-$B_2$/high-$D$ regime.
Even within the higher-degeneracy transition regime, P9 and P20 maintain overall high composition and sequence-descriptor similarity despite a 2.4-fold change in $B_2$ and a 1.9-fold change in $D$ between these points on the Pareto front.
\rev{Taken together, these observations suggest that there is substantial flexibility in choosing combinations of sequence features to achieve near-Pareto-optimal condensate properties.}

\rev{We next consider whether specific sequence characteristics, rather than overall sequence similarities, might control property changes across the Pareto front.
In this way, we find that a subset of sequence descriptors are most relevant and that the relative importance of each descriptor depends on the regime. }
In \figref{fig:features}c, we highlight the trends along the Pareto front of key sequence descriptors including sequence length, $N$; average sequence hydropathy, $\bar\lambda$; sequence hydropathy decoration, SHD; the absolute net charge per residue, $|\bar{q}|$; the fraction of charged residues (or the total charge), $\bar{q}_+ + \bar{q}_-$; and sequence charge decoration, SCD.
\rev{Data for other features that have previously been cited as relevant to IDP condensates~\cite{hazra2021biophysics,zheng2020hydropathy,statt2020model, mao2010net} are shown in Fig.~S9 in the Supplementary Materials.}
The highlighted descriptors display an array of behaviors when moving across the Pareto front.
$N$ is roughly sigmoidal with $-B_2$, starting at small $N$ in the high-$B_2$/high-$D$ regime (P1--P8) before gradually but noisily increasing to plateau at large $N$ in the low-$B_2$/low-$D$ regime (P27--P35).
By contrast, $\bar\lambda$ and SHD exhibit U-shaped behavior, first generally decreasing in the high-$B_2$/high-$D$ regime and then increasing once entering the transition region. 
Interestingly, the net charge, $|\bar{q}|$, of Pareto-optimal sequences is always close to zero, but there are qualitative differences in total charge, $\bar{q}_+ + \bar{q}_-$, and SCD in different regimes.
In particular, $\bar{q}_+ + \bar{q}_-$ starts at finite fractional charge at high-$B_2$/high-$D$ and initially increases before diminishing to zero in the transition and low-$B_2$/low-$D$ regimes; SCD displays similar behavior, but with the opposite sign.  

\rev{The collective observations from \figref{fig:features}c suggest two important guiding principles regarding the thermodynamics--dynamics tradeoff of protein condensates. 
First, the trends in charge-related features indicate that Pareto-optimal polypeptides are polyampholytic only in the high-$B_2$/high-$D$ regime.}
Polypeptides in this regime also tend to be shorter and less hydrophobic than other phase-separating IDPs generated during active learning.
Therefore, polyampholicity in combination with nonuniform charge distribution and weak hydrophobicity is key to achieving phase separation with short (and therefore fast-diffusing) chains.
\rev{Second, and by contrast, Pareto-optimal polypeptides are largely composed of uncharged, hydrophobic amino acids in the low-$B_2$/low-$D$ regime.}
In particular, sequences P27--P35 are predominantly composed of tyrosine, tryptophan, and phenylalanine, which have the highest hydrophobic character in the CG model and are also among the more massive amino acids. 
Greater hydrophobicity and longer chain lengths therefore result in both stronger inter-chain attraction and more sluggish diffusion in the low-$B_2$/low-$D$ regime.
\rev{It is notable that these insights are consistent with prior computational and experimental studies of protein-condensate rheology~\cite{hazra2021biophysics,alshareedah2023sequence,Rekhi2023.03.02.530853}.
Therefore, these principles may provide insights into general property trends resulting from sequence modifications, even though they stem from analyzing a specific set of Pareto-optimal polypeptides. }

\subsection*{Pareto-optimality is characterized by regime-dependent sequence determinants}
\label{sec:pareto-counterfactuals}

We perform a counterfactual analysis~\cite{wachter2017counterfactual,dandl2020multi,R:2023_Wellawatte_Perspective} of the Pareto-optimal IDPs to gain further insight into the sequence determinants of the heteropolymer thermodynamics--dynamics tradeoff. 
In this context, a ``counterfactual'' is an IDP sequence that bears strong sequence similarity to a Pareto-optimal sequence but exhibits distinctly different physical properties (\figref{fig:features}a).
Such counterfactuals can therefore be used to attribute Pareto-optimality to small variations in IDP sequences.
Here, we consider a polypeptide to be a counterfactual of a specific sequence on the Pareto front if \emph{(i)} the cosine similarity of their feature vectors is greater than 0.9 and \emph{(ii)} the counterfactual candidate is within a dimensionless distance of 0.15--0.3 away from the Pareto front; for this calculation, we use a Euclidean distance in the in the standard-normalized $B_2$--$D$ plane (see Supplementary Materials for details).
\rev{Requiring that the dimensionless distance is in the range 0.15--0.3 ensures that the properties of counterfactual sequences are statistically resolvable from those on the Pareto front while remaining in a similar regime of physical property behavior.}
Because each Pareto-optimal polypeptide may have several counterfactuals satisfying the aforementioned criteria, we examine the average feature differences,
\begin{equation}
  \label{eq:counterfactual}
    \langle \Delta x_k \rangle \equiv \frac{1}{n_{\mathcal{C}_{ij}}}\sum_{j=1}^{n_{\mathcal{C}_{ij}}}(x^{(\mathcal{P}_i)}_k-x^{(\mathcal{C}_{ij})}_k),
\end{equation}
where $k$ denotes a particular feature of a feature vector $\vec x$, $\mathcal{P}_i$ is the $i$th Pareto-optimal sequence with its value of $x_k$ indicated as $x_k^{(\mathcal{P}_i)}$, $\mathcal{C}_{ij}$ is the $j$th counterfactual of the $i$th Pareto-optimal sequence with its value of $x_k$ indicated as $x_k^{(\mathcal{C}_{ij})}$, and $n_{\mathcal{C}_{ij}}$ is the total number of counterfactuals of the $i$th Pareto-optimal sequence.
\rev{Plotting $\langle \Delta x_k  \rangle$ for all features versus $-B_2/V_0$ (\figref{fig:features}d and Fig.~S10 in the Supplementary Materials) allows us to ascertain whether specific sequence descriptors distinguish Pareto-optimal versus non-Pareto-optimal sequences.}

\rev{In the low-$B_2$/low-$D$ regime, we find that Pareto-optimal polypeptides possess several distinctive features relative to their counterfactual sequences.
In particular, Pareto-optimal sequences within this regime are generally longer, feature more hydrophobic residues, and have larger values of the sequence hydropathy decoration parameter, SHD, than their counterfactuals.}
Pareto-optimal sequences in this regime also tend to have fewer charged residues and a lower net charge than their counterfactuals.
Since the Pareto-optimal sequences in this regime are also neutral (\figref{fig:features}c), this implies that the counterfactual sequences possess a net charge despite having near zero values of the sequence charge decoration parameter, SCD (\figref{fig:features}c,d).
Our counterfactual analysis therefore indicates that hydrophobic interactions are the dominant determinants of Pareto-optimal sequences in the low-$B_2$/low-$D$ regime.

However, outside the low-$B_2$/low-$D$ regime, differences between counterfactuals and Pareto-optimal sequences are not as easily resolved. 
In the high-$B_2$/high-$D$ regime, no single sequence descriptor differs consistently between Pareto-optimal and counterfactual sequences.
Instead, variations in $B_2$ and $D$ manifest via subtle manipulations, such as exchanging residues with comparable hydrophobicity but different masses (Fig.~S10 in the Supplementary Materials).
Within the transition regime, there are statistically significant differences between features of Pareto-optimal polypeptides and counterfactuals, but the directions of such differences are not necessarily preserved across the Pareto front.
In particular, the hydrophobicity differences, $\langle \Delta\bar\lambda \rangle$, and total charge differences, $\langle \Delta (\bar{q}_++\bar{q}_-)\rangle$, fluctuate between positive and negative values within this regime (P9--P20 in \figref{fig:features}d).
This consistently enhanced variability across multiple features (see also Fig.~S10 in the Supplementary Materials) can be attributed to the greater degeneracy of sequence space near the Pareto front in the transition regime, as discussed in the previous section.
Overall, this counterfactual analysis indicates that the relationship between sequence features and Pareto optimality is nuanced, since relatively subtle changes from counterfactuals dictate behavior in the high-$B_2$/high-$D$ regime and myriad feature combinations exist for achieving near-optimality in the transition regime.

\section*{Discussion}

Using an active-learning approach, we have investigated the extent to which thermodynamic and dynamical properties are coupled in IDP condensates.
Our approach \rev{uniquely} combines high-throughput simulations of a coarse-grained model of IDPs, Bayesian active learning strategies for predicting physical properties from IDP sequences, and numerical optimization techniques to efficiently explore sequence space.
\rev{Overall, this approach confirms that a tradeoff between thermodynamic stability and internal condensate dynamics is a general, albeit tunable, feature of protein-based condensates.
Importantly, this work provide a robust assessment of condensate dynamics via a global analysis of IDP sequence space, rather than relying on systematic sequence mutations alone.}

\rev{The essential insight of our active-learning study is that heteromeric polypeptides, by virtue of their diverse physicochemical properties, tend to decouple the thermodynamic and dynamical properties of IDP condensates.}
In our study, the correlation between proxies for thermodynamic stability and condensed-phase dynamics is strongest in the case of \textit{homomeric} polypeptides.
In particular, both the thermodynamic stability and the condensed-phase dynamics of homomeric polypeptide condensates can be predicted with high accuracy on the basis of the second-virial coefficient and the chain length alone.
However, this correlation between thermodynamic and dynamic properties weakens when we consider \textit{heteromeric} IDP sequences, indicating that we can design sequences to tune the thermodynamic and dynamic properties of IDP condensates independently.
Consistent with this notion, we observe that the second-virial coefficient is not always sufficient for predicting phase separation of heteromeric IDP sequences.
Ultimately, by identifying a representative set of Pareto-optimal polypeptides---heteromeric sequences that outline a limiting tradeoff boundary with respect to the second-virial and diffusion coefficients---\rev{we find that the condensed-phase self-diffusion coefficient can be substantially increased relative to homomeric polypeptides with the same second-virial coefficient.}

\rev{Our systematic approach also provides insight into specific sequence descriptors that govern Pareto-optimal sequences.
However, we find that these sequence determinants are surprisingly nuanced and differ depending on the position of a sequence along the thermodynamics--dynamics Pareto front.} 
Whereas the extremal regions (high-$B_2$/high-$D$ and low-$B_2$/low-$D$) display consistent trends with respect to key sequence features, the transition regime is characterized by high sequence variability, which likely reflects greater degeneracy with respect to sequence space within this regime.
\rev{Chain length, hydrophobicity, and hydrophobic patterning are consistently important determinants of Pareto-optimality.
By contrast, total charge and charge patterning descriptors are primarily relevant in the high-$B_2$/high-$D$ regime that is dominated by polyampholytes, and enhancements to charge patterning tend to suppress polypeptide diffusion.}
Finally, our counterfactual analysis indicates that relatively minor sequence perturbations can result in large changes to both $B_2$ and $D$, particularly in the transition regime.
\rev{These findings suggest that different design principles are relevant in different regimes of physicochemical properties,} and they highlight the importance of considering multiple sequence features when optimizing polypeptide sequences to achieve specific physical properties.

We emphasize that our numerical results apply to single-component protein condensates and rely on the accuracy of the underlying coarse-grained IDP model, which lacks physicochemical features such as directional sidechain interactions, hydrogen bonds, and secondary structure.
\rev{These limitations of the underlying coarse-grained model influence the Pareto-optimal sequences, many of which look dissimilar from known naturally occurring IDPs, despite having overall low sequence complexity.
In this sense, our designed sequences should be interpreted as polypeptide-like heteropolymers with hydrophobic, electrostatic, and excluded-volume interactions as opposed to literal amino-acid sequences.}
Coarse-grained models are also known to predict dynamical properties with reduced fidelity compared to all-atom models with many more degrees of freedom~\cite{salvi2019solvent,R:2019_Rudzinski_Recent,R:2021_Dhamankar_Chemically}.
It is likely that a greater range of variability of dynamical properties would be predicted by a model that incorporates such features, especially if such modifications were to allow for more accurate descriptions of low-density condensed phases~\cite{wei2017phase}.
It is also likely that the range of variability of dynamic properties would be increased by examining IDP sequences with lengths greater than the upper limit of 50 used in this study.
\rev{Nonetheless, despite these considerations, we expect that our findings regarding the existence of a thermodynamics--dynamics tradeoff, as well as the relative importance of chain length, charge patterning, and hydrophobicity along the Pareto front, are robust with respect to alternative IDP models~\cite{joseph2021physics,garaizar2021salt,latham2019maximum,benayad2020simulation,tesei2021accurate,tesei2022improved} and actual polypeptide-like polymers.}
Experiments and higher-resolution simulations~\cite{salvi2019solvent,zheng2020molecular,galvanetto2022ultrafast,gruijs2022disease} will be needed to test such predictions.

A major strength of our approach is its ability to explore IDP sequence space efficiently and characterize relationships among disparate biophysical properties.
Crucially, the strategy of combining high-throughput simulations and Bayesian-guided optimization is independent of any limitations of the underlying coarse-grained model.
Moreover, we have demonstrated that this approach can efficiently converge the Pareto front for a pair of biophysical properties that depend on a complex sequence of computations.
We anticipate that this framework could be applied to other combinations of biophysical traits and may also be useful for studying multicomponent and multiphasic systems~\cite{jacobs2023theory,do2022engineering,chew2023physical}.
Our work therefore establishes a promising approach for engineering custom biomolecular condensates with tunable thermodynamic and dynamic properties.

\section*{Materials and Methods}

\subsection*{Model of intrinsically disordered proteins}
\label{sec:model}

The coarse-grained (CG) model developed by Regy et al.~\cite{regy2021improved} represents IDP sequences, comprising 20 amino-acid types, in implicit solvent.
The force field consists of a harmonic bonded potential as well as short-range van der Waals (vdW) and long-range electrostatic (el) nonbonded interactions,
\begin{equation} \label{eqn:1}
  U_{\text{tot}} = \sum_i k_{\text{b}}(r_{i,i+1}-b_0)^2+\sum_{i,j}\phi^{\text{vdw}}(r_{ij})+\sum_{i,j}\phi^{\text{el}}(r_{ij}),
\end{equation}
where $r_{i,i+1}$ is the distance between bonded residues $i$ and $i+1$, the force constant is $k_{\text{b}}=\SI{10}{kcal/(mol\cdot \AA^2)}$, and the equilibrium bond length is $b_0=3.82$ \AA.
The vdW interaction takes the Ashbaugh--Hatch functional form,
\begin{equation}  \label{eqn:2}
  \phi^{\text{vdW}}(r_{ij})=\begin{cases}
  \phi^{\text{LJ}}(r_{ij})+(1-\lambda_{ij})\epsilon &\quad r_{ij} \leq 2^{1/6}\sigma_{ij}\\
  \lambda_{ij}\phi^{\text{LJ}}(r_{ij}) &\quad  r_{ij} > 2^{1/6}\sigma_{ij}\\
  \end{cases}
\end{equation}
where $r_{ij}$ is the distance between nonbonded residues $i$ and $j$, and $\phi^{\text{LJ}}(r_{ij}) = 4\epsilon[(\sigma_{ij}/r_{ij})^{12} - (\sigma_{ij}/r_{ij})^6]$ is the Lennard--Jones potential with $\epsilon=0.2$ kcal/mol.
The interaction strength, $\lambda_{ij} = (\lambda_i + \lambda_j)/2$, and distance, $\sigma_{ij} = (\sigma_i + \sigma_j)/2$, parameters are determined from the hydropathy scaling factors, $\lambda_i$, and vdW diameters, $\sigma_i$, of the residues, respectively~\cite{regy2021improved}.
Lastly, screened electrostatic interactions take the form
\begin{equation} \label{eqn:4}
  \phi^{\text{el}}(r_{ij}) = \frac{q_i q_j}{4\pi Dr_{ij}} e^{-\kappa r_{ij}},
\end{equation}
where $q_i$ is the charge of residue $i$, $D=80$ is the dielectric constant of the solvent, and $\kappa=10$ \text{\AA} is the Debye screening length.

\subsection*{Physical property calculations}
\label{sec:property-calculations}

\noindent \textit{Second-virial coefficient.}
Second-virial coefficients are determined by calculating the potential of mean force, $u(r)$, as a function of the center-of-mass (COM) distance, $r$, between an isolated pair of chains.
These calculations are carried out using the adaptive biasing force (ABF) method~\cite{comer2015adaptive} implemented in the COLVARS package in LAMMPS~\cite{thompson2022lammps}.
A Langevin thermostat is applied to maintain a constant temperature of \SI{300}{K}.
Two molecules are placed in a cubic simulation box with dimensions $300\times300\times300$ {\AA}.
ABF simulations are performed as a function of $r$, which varies from \SI{1}{~\angstrom} to \SI{100}{~\angstrom}.
Force statistics are stored in bins of width \SI{0.2}{\angstrom}.
The biasing force is applied after 1000 samples are collected in each bin, after which a final production run is performed for \SI{5}{\micro\second} with a timestep of \SI{10}{\fs}.
The second-virial coefficient is then obtained from \eqref{eq:B2}.
Thirty independent simulations are performed for each free energy calculation to determine the statistical error.

\vskip1ex\noindent \textit{Condensed-phase density.}
Considering the high computational costs of slab simulations, we developed an alternative method to determine whether a CG sequence undergoes phase separation at 300 K.
To this end, we look for evidence of a van der Waal's loop in the equation of state (EOS) of a small system.
Specifically, we compute the pressure using canonical-ensemble simulations of 100 chains with periodic boundary conditions at densities ranging from \SI{0.2}{g/ml}--\SI{1.2}{g/ml}.
A Langevin thermostat is applied to maintain a constant temperature of \SI{300}{K}.
Simulations are carried out for \SI{100}{\ns} with a timestep of \SI{10}{\fs}.
\rev{We assume a near-zero coexistence pressure to minimize statistical and fitting errors,} so that phase separation can be identified by the existence of a negative pressure at a finite density.
\rev{This approximation is justified by the fact that the coexisting dilute phase is characterized by a low polymer density and is nearly ideal, implying that the coexistence pressure is also close to zero.}
In cases where negative pressures are observed, we fit a cubic spline to the EOS to determine the condensed-phase density, $\rho_{\text{c}}$, defined as the highest density at which the pressure crosses zero.
The pressure values are bootstrapped with replacement 50 times to determine the average condensate densities and their standard errors.

\vskip1ex\noindent \textit{Self-diffusion coefficient.}
To determine the self-diffusion coefficient of a chain within the condensed phase, canonical-ensemble simulations of 100 chains are performed at the condensed-phase density, $\rho_{\text{c}}$, with periodic boundary conditions.
The long-time behavior of the mean-squared displacement is then used to compute the self-diffusion coefficient,
\begin{equation}
  D = \lim_{\Delta t\to\infty} {\frac{1}{6\Delta t} \langle |\vec{r}_{t+\Delta t} - \vec{r}_t|^2 \rangle},
\end{equation}
where $\vec{r}$ is the position of the COM of a tagged chain.
The temperature is maintained at \SI{300}{K} using a Langevin thermostat with a damping frequency of \SI{1}{\ps^{-1}}.
Simulations are carried out for \SI{100}{\ns} with a timestep of \SI{10}{\fs}.
For each diffusion-coefficient calculation, the statistical error is quantified using 30 independent simulations.

\subsection*{Design of novel intrinsically disordered proteins}
\label{sec:ml}

\noindent \textit{Overview of framework.}
We deploy an active learning~\cite{R:2018_Smith_Less,R:2016_Zuluaga_pare} framework based on Bayesian optimization~\cite{shahriari2015taking} to identify IDP sequences that define a Pareto front of the properties $B_2$ and $D$. The active learning framework utilizes coarse-grained simulations to generate data for $B_2$ as well as pressure-density data and $D$, as appropriate.
These data are used to train three different machine-learning (ML) models: two Gaussian process regressors~\cite{rasmussen2006gaussian} (GPR) and one random forest (RF) binary classifier~\cite{biau2016random}.
These ML models are used to make surrogate predictions during the optimization step of active learning, during which new IDP sequences are selected for simulation.
Sequence optimization is carried out using a genetic algorithm, as described below.
Following optimization, simulations are performed to generate new data and begin the next iteration of active learning.  

\vskip1ex\noindent \textit{Machine-learning models.} 
GPR models are trained to predict $B_2$ and $D$, while a RF model is trained to predict whether a given IDP sequence will undergo phase separation. 
For a given sequence $i$, the input for all ML models is $\vec{x}^{(i)}$, size-explicit augmented fingerprint as a feature vector~\cite{R:2022_Patel_Featurization}.
In particular, the feature vector is 30-dimensional, consisting of 20 features reflecting the composition of the amino acids and ten sequence-level chain descriptors.
The elements of the feature vector are all normalized via linear transformation methods to be unit-scale.
Additional details regarding the featurization, training, and hyperparameter optimization for all ML models are provided in the Supplementary Materials.
Although model accuracy is not a primary objective, the three trained ML models show high accuracy in predicting the aforementioned properties (see Fig.~S6 in the Supplementary Materials) at the conclusion of active learning. 
The coefficients of determination  $R^2$ of the two GPR models are as high as $0.975\pm0.015$ and $0.970\pm0.015$, and the accuracy of the RF classifier is 0.96. 

\vskip1ex\noindent \textit{Sequence optimization.} 
Different fitness functions were used for sequence design by a genetic algorithm depending on the active-learning iteration.
Denoting the iteration number by the index $k$, the fitness function  $f(\vec{x}^{(i)})$~is
\begin{equation}
  \label{eqn:fitness}
  f(\vec{x}^{(i)})\! = \!\begin{cases}
  -B_2(\vec{x}^{(i)}) & \text{if } k = 0 \\
  \text{EHVI}(\vec{x}^{(i)})\times\text{RF}(\vec{x}^{(i)})\times\alpha(\vec{x}^{(i)})\!\!\!\! & \text{if } 0 < k < 8 \\
  -\tilde{B}_2(\vec{x}^{(i)}) + \tilde{D}(\vec{x}^{(i)}) & \text{if } k = 8,
  \end{cases}
\end{equation}
where EHVI($\vec{x}^{(i)}$) is the expected hypervolume improvement acquisition function,
RF($\vec{x}^{(i)}$) is the output label of the random forest classifier (evaluating to 1 if phase separation is predicted and to 0 otherwise), $\alpha(\vec{x}^{(i)})$ is a scaling function that biases against generation of sequences that have high similarity to previously simulated or proposed sequences, and $\tilde{A}$ indicates a standard-normalized transformation of the property $A$ based on the data acquired up to the given iteration. 
For each iteration, 96 unique sequences are generated. 
In {\eqref{eqn:fitness}}, we compute the EHVI~\cite{R:2019_Yang_Multi} for a given sequence $\vec{x}^*$,
\begin{multline}
  \label{eqn:8}
  \text{EHVI}(\vec{x}^*,\mathcal{P}') =  \sum_{i=1}^{n+1} \Psi(\tilde{D}^{(i)},\tilde{D}^{(i)},\tilde{D}^*,\sigma^*_{\tilde{D}}) \\
  \times \biggl[ (\tilde{B}_2^{(i)}-\tilde{B}_2^{(i-1)})\Phi \Bigl(\frac{\tilde{B}_2^* -\tilde{B}_2^{(i)} }{\sigma^*_{\tilde{B}_2}} \Bigr) \\
    \qquad + \Bigl( \Psi(-\tilde{B}_2^{(i-1)},-\tilde{B}_2^{(i-1)}-\tilde{B}_2^*,\sigma^*_{\tilde{B}_2} ) \\
    \qquad - \Psi(-\tilde{B}_2^{(i-1)},-\tilde{B}_2^{(i)},  -\tilde{B}_2^*,\sigma^*_{\tilde{B}_2} ) \Bigr) \biggr],
\end{multline}
where $\tilde{B}_2^*$ and $\tilde{D}^*$ are predicted values of the second-virial and diffusion coefficients, respectively, obtained from the GPR models; $\sigma^*_{\tilde{B}_2}$ and $\sigma^*_{\tilde{D}}$ are corresponding uncertainty estimates from the GPR models; $\mathcal{P}' = \{\vec{x}^{(0)}, \vec{x}^{(1)},\ldots,\vec{x}^{(n)}, \vec{x}^{(n+1)} \}$ represents the current $n$-point approximation of the Pareto front, augmented with additional reference points $\vec{x}^{(0)} = (X, -\infty)^{\top}$ and $\vec{x}^{(0)} = (-\infty,X)^{\top}$ to facilitate the hypervolume calculation; $\Phi(s)$ denotes the cumulative probability distribution function for a standard normal distribution of $s$; and $\Psi$ is a function defined as 
\begin{equation}
  \Psi(a,b,\mu,\sigma) = \int_{-\infty}^b \frac{(a-z)}{\sigma} \frac{1}{\sqrt{2\pi}} e^{-\frac{(z-\mu)^2}{2\sigma^2}} dz.
\end{equation}
The similarity penalty $\alpha(\vec{x^{(j)}})$ is defined to be
\begin{equation}
  \label{eqn:9}
  \alpha(\vec{x}^{(j)}) = \frac{j-1}{\sum_{k=1}^{j-1} \left\{\frac{1}{2}[1-\vec{x}^{(j)} \cdot \vec{x}^{(k)}/(|\vec{x}^{(j)}||\vec{x}^{(k)}|)]\right\}^{-1}},
\end{equation}
where $j=2,3,\ldots,96$ are indices of candidate sequences generated in a given iteration.
When a sequence is highly similar to one or more previously generated sequences, $\alpha$ tends to zero.
For $j=1$, we take $\alpha = 1$ (i.e., no penalty is applied).

The active learning process is seeded with data obtained from simulations of 1,266 disordered sequences reported in DisProt~\cite{hatos2020disprot}.
We select and simulate all sequences with sequence lengths between 20 and 50 residues (inclusive), taking care to remove duplicated samples. 
Since none of these sequences are determined to phase separate, we generate new sequences by maximizing a fitness function that involves only $B_2$ in iteration 0.
The resulting data enable the training of ML models for $B_2$, $D$, and phase separation.
Then in subsequent iterations, we maximize a fitness function based principally on the expected hypervolume improvement (EHVI) to identify polypeptides that outline a Pareto front of $-B_2$ versus $D$.
To confirm the convergence of the Pareto front, a final exploitation-only round of optimization is performed by maximizing a simple linear function of the standard-normalized properties $-\tilde B_2$ and $\tilde D$.

For each iteration, 96 ``child'' sequences are successively generated that aim to maximize the relevant fitness function in \eqref{eqn:fitness}.
To facilitate convergence towards selecting each child sequence, 96 independent trials of sequence optimization are executed in parallel, and only the sequence with the best fitness score is chosen for explicit simulation; overall, this means that $96\times96$ sequence optimizations are performed in each iteration. 

In iteration 0, the initial ``parent'' sequences are DisProt sequences with negative $B_2$. 
In all subsequent iterations, the parent sequences are Pareto-optimal sequences identified from the previous iteration.
Candidate child sequences are derived from the parent sequences via a combination of crossover and mutation moves.
In addition, ``deletion'' and ``growth'' moves are performed, which either remove a portion of the current sequence or add a portion of the current sequence within itself.
These moves are executed \emph{independently} in the order of crossover, mutation, deletion and growth with probabilities of 0.5, 0.8, 0.2, and 0.5, respectively.
These moves are performed for 100 steps, which is sufficient for the genetic algorithm to converge to a local maximum of the fitness function (Fig.~S11 in the Supplementary Materials).
Additional details can be found in the Supplementary Materials.

\nocite{zhang2015reliable}

\section*{Acknowledgments}

The authors acknowledge assistance from Princeton Research Computing at Princeton University, which is a consortium led by the Princeton Institute for Computational Science and Engineering (PICSciE) and the Princeton University Office of Information Technology's Research Computing. 
\paragraph*{Funding:}
This work was a grant from the Princeton Biomolecular Condensate Program to M.A.W. and W.M.J.
\paragraph*{Author contributions:} 
Conceptualization: M.A.W., W.M.J.;
Methodology: Y.A., M.A.W., W.M.J.;
Investigation: Y.A., M.A.W., W.M.J.;
Visualization: Y.A., M.A.W., W.M.J.;
Supervision: M.A.W., W.M.J.;
Writing (original draft): Y.A., M.A.W., W.M.J.;
Writing (review and editing): Y.A., M.A.W., W.M.J.
\paragraph*{Competing interests:} The authors declare that they have no competing interests.
\paragraph*{Data and materials availability:} All data needed to evaluate the conclusions of the paper are available in the main text or in the supplementary materials. Full datasets compiled from this work are provided in Ref. \cite{R:2023_Webb_Thermodynamic}. \rev{Demonstrative Jupyter notebooks for essential calculations are also provided at https://github.com/webbtheosim/condensate-public.}

\end{document}


\onecolumngrid

\section*{Supplementary Information for ``Active learning of the thermodynamics--dynamics tradeoff in protein condensates''}\vskip-2ex
\author{Yaxin An}
\email{Present address: Department of Chemical Engineering, Louisiana State University, Baton Rouge, LA 70803 USA}
\affiliation{%
  Department of Chemical and Biological Engineering, Princeton University, Princeton, NJ 08544 USA\\
}%
\affiliation{%
  Department of Chemistry, Princeton University, Princeton, NJ 08544 USA
}%
\author{Michael A. Webb}%
 \email{mawebb@princeton.edu}
\affiliation{%
  Department of Chemical and Biological Engineering, Princeton University, Princeton, NJ 08544 USA\\
}%
\author{William M. Jacobs}%
 \email{wjacobs@princeton.edu}
\affiliation{%
  Department of Chemistry, Princeton University, Princeton, NJ 08544 USA
}%

\date{\today}

\onecolumngrid
\maketitle

\onecolumngrid

\section{Extended methods}
\label{sec:extended_methods}

\subsection{Featurization of protein sequences}
\label{sec:features}

A 30-dimensional feature vector $\vec{x}$ is used as a numerical encoding of a polypeptide sequence and its chemical characteristics. To construct this feature vector, 30 different characteristics of each protein sequence are considered, including 
\begin{itemize}
    \item the composition of each amino acid in the sequence, $c_a$, where $a$ corresponds to one of the amino acids (A, C, D, E, F, G, H, I, K, L, M, N, P, Q, R, S, T, V, W, and Y) and $\sum_a c_a = 1$;
    \item the sequence length (i.e., number of amino acids), $N$;
    \item the net charge per residue, $|\bar q|$,
      \begin{equation*}
        |\bar q| \equiv \frac{1}{N}\left|\sum_{i=1}^N q_i\right| ;
      \end{equation*}
    \item the sequence charge decoration, SCD,
      \begin{equation*}
        \text{SCD} \equiv \frac{1}{N}\sum_{i=1}^N\sum_{j=i+1}^{N} q_iq_j(j-i)^{1/2};
      \end{equation*}
    \item the average hydrophobicity per residue, $\bar\lambda$,
      \begin{equation*}
        \bar\lambda \equiv \frac{1}{N} \sum_{i=1}^N \lambda_i ;
      \end{equation*}
    \item the sequence hydropathy decoration, SHD,
      \begin{equation*}
        \text{SHD} \equiv \frac{1}{N} \sum_{i=1}^N\sum_{j=i+1}^N(\lambda_i+\lambda_j)(j-i)^{-1};
      \end{equation*}
    \item a mean-field prediction of the second-virial coefficient, $B_2^{({\text{MF}})}$,
      \begin{equation*}
        B_2^{({\text{MF}})} \equiv \sum_i^N\sum_j^Nb_{2,ij},
      \end{equation*}
      where $b_{2,ij} \equiv 2\pi \int_0^\infty dr\, r^2 \left[1 - e^{-\beta u_{ij}(r)}\right]$
      is the second-virial coefficient between monomers $i$ and $j$, which are assumed to be on different chains, and $u_{ij}(r)$ is the pair potential between monomers $i$ and $j$;
    \item the fraction of positively charged residues, $\bar q_{+}$;
    \item the fraction of negatively charged residues, $\bar q_{-}$;
    \item the Shannon entropy, $S$,
      \begin{equation*}
        S \equiv -\sum_a c_a \log c_a;
      \end{equation*}
    \item and the average molar mass of a residue, $\bar M$,
      \begin{equation*}
        \bar M \equiv \frac{1}{N}\sum_{i=1}^N M_i,
      \end{equation*}
      where $M_i$ is the molecular weight of the $i$th amino acid in the sequence.
\end{itemize}
For the purpose of using these sequence characteristics as features in machine learning and optimization, it is useful to transform the variables such that they have similar magnitudes. 
Therefore, we apply normalization techniques to each of the last ten sequence characteristics (i.e., all features except the $\{c_a\}$, which are already on a unit scale).
With the exception of $N$ and $S$, standard normalization is used for each feature:
\begin{equation}
  \tilde{A} = \frac{A - \mu_A}{\sigma_A},
\end{equation}
where $A$ is a given sequence characteristic with an average of $\mu_A$ and a standard deviation of $\sigma_A$ estimated over the sequences that comprise the current dataset $\mathcal{D}$ for training the machine-learning models.
Min-max normalization is employed for $N$,
\begin{equation}
  \tilde{N} = \frac{N-\min_\mathcal{D}{N} }{\max_\mathcal{D}{N}-\min_\mathcal{D}{N}},
\end{equation}
and normalization by the maximum is employed for $S$,
\begin{equation}
  \tilde{S} = \frac{S}{\max_\mathcal{D}{S}} -1.
\end{equation}
We note that similar feature vectors have demonstrated excellent predictive capabilities relating to both structural and dynamic properties of intrinsically disordered protein chains~(58,59).

\subsection{Gaussian process regression models}
\label{sec:gpr}

Gaussian process regression (GPR) is used to estimate expected values and uncertainties for the second-virial coefficient, $B_2$, and the condensed-phase self-diffusion coefficient, $D$, as a function of the feature vector, $\vec{x}$.
GPR is particularly used for its native uncertainty estimates, which are central to the computation of expected hypervolume improvement (EHVI) acquisition function described in the main text, \eqref{eqn:kernel}.
Separate GPR models are constructed for $B_2$ and $D$. 
Covariances modeled by the Gaussian processes are calculated using a Mat\'{e}rn kernel ($\nu = {3/2}$) with added noise,
\begin{equation}\label{eqn:kernel}
k(\vec{x},\vec{x}') = \sigma^2  \Biggl[ \frac{1}{\Gamma({3/2}) 2^{{1/2}}} \times \left(\frac{\sqrt{3}}{l} || \vec{x} - \vec{x}' ||_2 \right)^{3/2} K_{3/2}  \left( \frac{\sqrt{3}}{l} || \vec{x} - \vec{x}' ||_2 \right) \Biggr] + \sigma_n^2.
\end{equation}
In \eqref{eqn:kernel}, $\sigma$, $l$, and $\sigma_n$ are treated as adjustable hyperparameters, $\Gamma (\cdot)$ denotes the gamma function, and $K_{3/2}$ is a modified Bessel function. Hyperparamater optimization is predicated on minimization of the mean-squared errors. 
Five-fold cross-validation is employed on the training set to mitigate overfitting and obtain optimal hyperparameters. 
In particular, five sets of optimized hyperparameters are produced (one for each fold), and these sets are then averaged to yield a final set of parameters. Using this final set, a GPR model is trained over the entirety of the data acquired up to that point for the purpose of surrogate modeling during sequence optimization.
The performance of the GPR models at the conclusion of active learning for an 80/20 train/test split is shown in \figref{fig:s6}a,b and illustrates good predictive ability.

\subsection{Random forest classifier construction}
\label{sec:rf}

A random forest (RF) classifier is used to predict whether phase separation is expected to occur as a function of the feature vector, $\vec{x}$.
Here, the RF classifier consists of 100 decision trees.
The best split in decision trees is determined by the Gini impurity. The minimum number of samples required to split an internal node is set at two, and the minimum number of samples required to be a leaf node is one. 
Five-fold cross-validation is employed on the training set to mitigate overfitting and to obtain optimal hyperparameters in the same fashion as described in SI~\secref{sec:gpr}.
The performance of the RF classifier at the conclusion of active learning for an 80/20 train/test split is shown as a confusion matrix in \figref{fig:s6}c.

\subsection{Sequence optimization and genetic algorithm}
\label{sec:genetic-algorithm}

Sequence optimization to identify candidate proteins with high EHVI is facilitated using a genetic algorithm.
In addition to conventional mutation and crossover moves, we also employ ``deletion'' and ``growth'' moves to improve diversity of proposed sequences.
All moves are constrained as necessary to produce sequences with $20\leq N \leq 50$. In all discussion, sequences of $N$ residues are index from 1 to $N$. In each iteration, batches of 96 candidate sequences are produced as described below. 

\begin{itemize}
\item \emph{Step 1}: A set of $n$ possible parent sequences sorted by their fitness from high to low. 
In Iteration 0, the parent sequences are sequences from DisProt that exhibit negative $B_2$.
At the beginning of every other iteration, the parent sequences correspond to those sequences that define the current approximation to the Pareto front.   

\item \emph{Step 2}: To generate ``child" sequences, a pair of sequences are randomly selected from the top 30\% of parent sequences. These sequences then undergo a series of crossover, mutation, growth and deletion moves:
  \begin{itemize}
    \item \textit{Crossover}: Suppose the pair of selected sequences have a length of $N$ and $M$. First, two indices are selected $s_1$ and $s_2$ both from the set $\in[1,\min(\{N,M\})$ with uniform probability. Then, another index is selected $s_3 \in [1,\max(\{N,M\})-(s_2-s_1)]$. Finally, the sub-sequence with the indices $\in[s_1,s_2]$ from the smaller sequence is exchanged with the sub-sequence with indices $\in[s_3,s_3+(s_2-s)1]$ from the larger sequence, accounting for changes to bond connectivity as needed.
            
    \item \textit{Deletion}: For a given sequence of length $N$, the length of a sub-sequence for deletion $l_\text{del}\in [0,N-20]$ is selected with uniform random probability. Next, an index $s$ is chosen from the set $[0,N-l_\text{del}]$. Then, the amino acids with indices in the sequence $\in$ $[s,s+l_{\text{del}}-1]$ are removed from the chain. If necessary, bonds are established between the amino acids with indices of $s-1$ and $s+l_\text{del}$. 
    
    \item \textit{Growth}: For a given sequence, the length of a sub-sequence for addition $l_\text{gro} \in [0,50-N]$ is selected with uniform random probability. Next, a random position $s$ in the sequence is selected with uniform probability. Then, the sub-sequence with indices $\in[1,l_\text{gro}]$ is replicated and inserted at $s$. As necessary, the bond between amino acids at $s$ and $s+1$ is eliminated, the amino acid at $s$ is bonded to the first amino acid in the replicated sequence, and the amino acid formerly at $s+1$ is bonded the last amino acid in the replicated sequence.

    \item \textit{Mutation}: For a given sequence of length $N$, an index $s \in [1,N]$ is selected with uniform random probability. Then, the amino acid at position $s$ is exchanged for another amino acid, which is selected from the pool of twenty amino acids with uniform random probability. 
  \end{itemize}
The probabilities for crossover, deletion, growth, and mutation are 0.5, 0.2, 0.5, and 0.5, respectively.  
The moves are executed independently in the order of crossover, deletion, growth, and mutation.
This procedure generates two new ``child'' sequences.

\item \emph{Step 3}: Repeat \emph{Step 2} until  $\round{0.7n}$ new sequences have been produced. These newly generated child sequences are combined with the top 30\% of the parent sequences from the prior generation to comprise a new set of $n$ parent sequences, which represent a new generation. 

\item \emph{Step 4}: The fitness function is evaluated over the new generation of prospective parent sequences. The sequences are ranked by fitness.  

\item \emph{Step 5}: Repeat starting from Step 2 until 100 generations have been produced.

\item \emph{Step 6:} The sequence with the best fitness at the conclusion of all generations is identified and proposed as a candidate for simulation as part of the active learning iteration. 

\emph{Steps 1-6} above results in one candidate sequence. To facilitate convergence towards an optimal sequence,  \emph{Steps 1-6} are executed 96 times in parallel (see \figref{fig:convergence}) without communication. The resulting 96 sequences, which differ by the stochastic nature of the genetic algorithm,  are ranked by their fitness, and the one with the best overall fitness is marked for characterization by simulation.  This overall workflow is then executed again (going back to \emph{Step 1}) 95 additional times to obtain a total of 96 distinct sequences for characterization by simulation. 
Note that the fitness function changes based on the active learning iteration and the prior history of proposed sequences (see Eqn. 7 of the main text).
Although a genetic algorithm was used for this task, any reasonable optimization algorithm should be able to achieve similar results.
In addition, different move probabilities or move sets may facilitate better convergence but were not explored here.
\end{itemize}

To characterize the stochasticity of the sequences produced by the genetic algorithm (GA), we perform five replicate rounds of GA to generate alternative sequences for the final ``exploitation'' step of our active learning approach (see Figure 2c in the main text).
The predicted $B_2$ and $D$ values of these polypeptides are shown in \figref{fig:ga_replica}, and representative sequences are listed in Tables~\ref{tab:GA_b1}--\ref{tab:GA_b4}.
In \figref{fig:ga_replica}, we consider four distinct regions along the Pareto front as examples for detailed comparisons.
Within each region, the sequences generated by different GA replicates are predicted to have similar $B_2$ and $D$ values and thus are predicted to lie in close proximity to the Pareto front.
However, as shown in Tables~\ref{tab:GA_b1}--\ref{tab:GA_b4}, the sequences generated by different GA replicates differ substantially from one another due to the stochastic nature of the GA optimization approach.
Nevertheless, in this exercise, we do not identify any sequences that notably improve upon our previously identified Pareto front.
This suggests that our stochastic optimization has converged to a representative boundary in terms of the thermodynamics-dynamics tradeoff, although precise sequences would likely differ were the study to be conducted again.

\subsection{Determination of counterfactuals}
\label{sec:dimensionless-distance}

The task of determining a counterfactual for a sequence requires identifying another sequence that possesses overall similar characteristics but exhibits a different classification.
In the context of this study, the classifications correspond to \emph{Pareto-optimal} versus \emph{near-Pareto-optimal}.
To treat our two objective properties ($D$ and $B_2$) on equal footing, we consider the classification of near-Pareto-optimal to be based on a distance determined according to standard-normalized variables $\tilde{D}$ and $\tilde{B}_2$.
Each sequence $i$ can then be represented as a coordinate in the  standard-normalized $B_2$--$D$ plane as $\vec{z}^{(i)} = (\tilde{B}_2^{(i)},\tilde{D}^{(i)})^\intercal$.
To account for statistical uncertainties associated with the quantities obtained from simulations, we associated near-Pareto-optimal with distances $d\in[0.15,0.3]$, which are measured from a given coordinate to a piecewise-linear approximation to the Pareto front.
In particular, we first calculate three sets of 35 distances ($\{d_{v,i}\}$, $\{d_{1,i}\}$ and $\{d_{2,i}\}$) associated with our 35 Pareto-optimal sequences with coordinates $\vec{z}^{(\text{P}_1)}, \dots,\vec{z}^{(\text{P}_{35})}$ as below:
\begin{equation}
   d_{v,i} = {\frac{(\vec{z}^{(k)} - \vec{z}^{(\text{P}_i)}) \times (\vec{z}^{(\text{P}_i)} - \vec{z}^{(\text{P}_{i+1})})}{|(\vec{z}^{(\text{P}_i)} - \vec{z}^{(\text{P}_{i+1})}|}} ,
\end{equation}

\begin{equation}
    d_{1,i} = |\vec{z}^{(k)} - \vec{z}^{(\text{P}_i)}|,
\end{equation}
and
\begin{equation}
    d_{2,i} = |\vec{z}^{(k)} - \vec{z}^{(\text{P}_{i+1})}|,
\end{equation}
where $d_{v,i}$ is the distance between the point $\vec{z}^{(k)}$ to the piecewise lines determined by two adjacent Pareto-optimal points $\vec{z}^{(\text{P}_i)}$ and $\vec{z}^{(\text{P}_{i+1})}$, $d_{1,i}$ is the distance between $\vec{z}^{(k)}$ and $\vec{z}^{(\text{P}_i)}$, and $d_{2,i}$ is the distance between $\vec{z}^{(k)}$ and $\vec{z}^{(\text{P}_{i+1})}$; for the above $i\in[1,34]$.  

A series of distances $d_k^{(i)}$ for $\vec{z}^{(k)}$  are then determined as
\begin{equation}
    d_k^{(i)} = \begin{cases}
        d_{v,i}  & \text{if $\cos \theta_{1,i} > 0$, and $\cos \theta_{2,i} < 0$} \\
        d_{1,i}  & \text{if $\cos \theta_{1,i} < 0$, and $\cos \theta_{2,i} < 0$} \\
        d_{2,i} & \text{if $\cos \theta_{1,i} > 0$, and $\cos \theta_{2,i} > 0$},
   \end{cases}
\end{equation}
where $\theta_{1,i}$ is the angle between $(\vec{z}^{(k)} -  \vec{z}^{(\text{P}_{i+1})})$ and $(\vec{z}^{(\text{P}_i)} - \vec{z}^{(\text{P}_{i+1})})$, and $\theta_{2,i}$ is the angle between  $(\vec{z}^{(k)} -  \vec{z}^{(\text{P}_{i})})$ and $(\vec{z}^{(\text{P}_i)} - \vec{z}^{(\text{P}_{i+1})})$:
\begin{equation}
  \cos\theta_{1,i} = \frac{(\vec{z}^{(k)} -  \vec{z}^{(\text{P}_{i+1})}) \cdot (\vec{z}^{(\text{P}_i)} - \vec{z}^{(\text{P}_{i+1})})}{|(\vec{z}^{(k)} - \vec{z}^{(\text{P}_{i+1})})| |(\vec{z}^{(\text{P}_i)} - \vec{z}^{(\text{P}_{i+1})})|},
\end{equation}
\begin{equation}
  \cos\theta_{1,i} = \frac{(\vec{z}^{(k)} -  \vec{z}^{(\text{P}_{i})}) \cdot (\vec{z}^{(\text{P}_{i})} - \vec{z}^{(\text{P}_{i+1})})}{|(\vec{z}^{(k)} - \vec{z}^{(\text{P}_{i})})| |(\vec{z}^{(\text{P}_i)} - \vec{z}^{(\text{P}_{i+1})})|}.
\end{equation}
Finally, the shortest distance of $z^{(k)}$ to the Pareto front is set as 
\begin{equation}
    d_k = \min (\{d_k^{(i)} \}). 
\end{equation}

\section{Supplementary Figures and Tables}

\begin{figure}[ht]
\centering
    \includegraphics{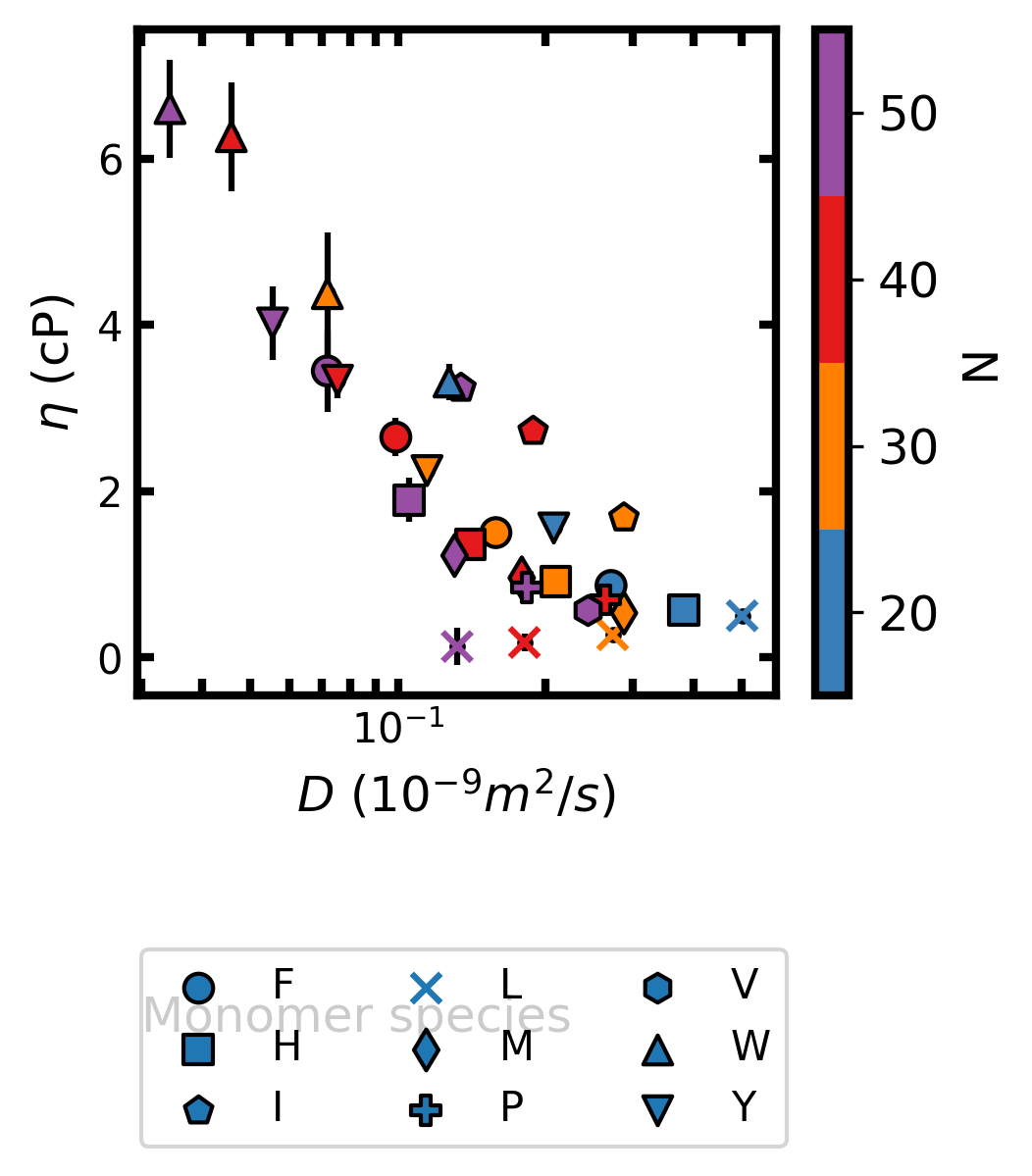}
    \caption{\textbf{Relationship of viscosity and self-diffusivity in the condensed-phase of homomeric polypeptides.} The viscosity, $\eta$, is anticorrelated with self-diffusivity, $D$, although the former possesses larger statistical uncertainties. The Pearson correlation coefficient between $\eta$ and $log(D)$ is $-0.91$. For all systems, quantities are measured in the canonical ensemble at the estimated coexistence density of the condensed phase as determined by the approximate EOS method (see main manuscript). The viscosity is determined using the Green--Kubo formalism~(85). Error bars represent the standard error of the mean.}
    \label{fig:s1}
\end{figure}

\begin{figure*}
    \includegraphics[width=0.95\linewidth]{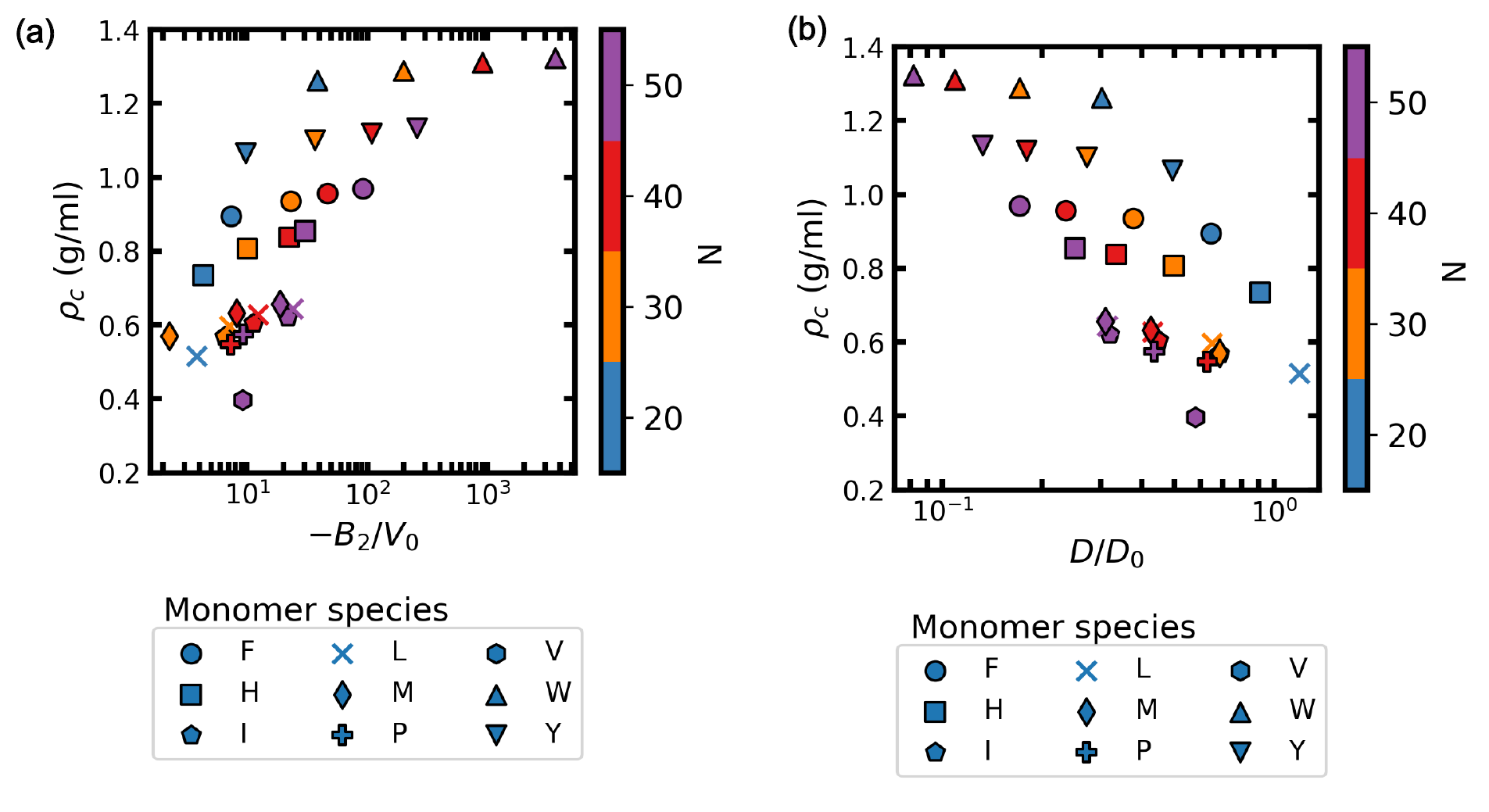}
    \caption{\textbf{Correlation of condensed-phase densities with target properties in phase-separating homomeric polypeptides.} Correlation plots for condensed-phase density, $\rho_{\text{c}}$ with \textbf{(a)} the dimensionless second-virial coefficient, $-B_2/V_0$, and \textbf{(b)} the dimensionless self-diffusion coefficient in the condensed-phase, $D/D_0$. The Pearson correlation coefficients between $\rho_{\text{c}}$ and $log(-B_2/V_0)$, and between $\rho_{text{c}}$ and $log(D/D_0)$ are $0.80$ and $-0.77$, respectively. Standard errors are comparable to symbol sizes.}
    \label{fig:s2}
\end{figure*}

\begin{figure*}
    \includegraphics[width=0.95\linewidth]{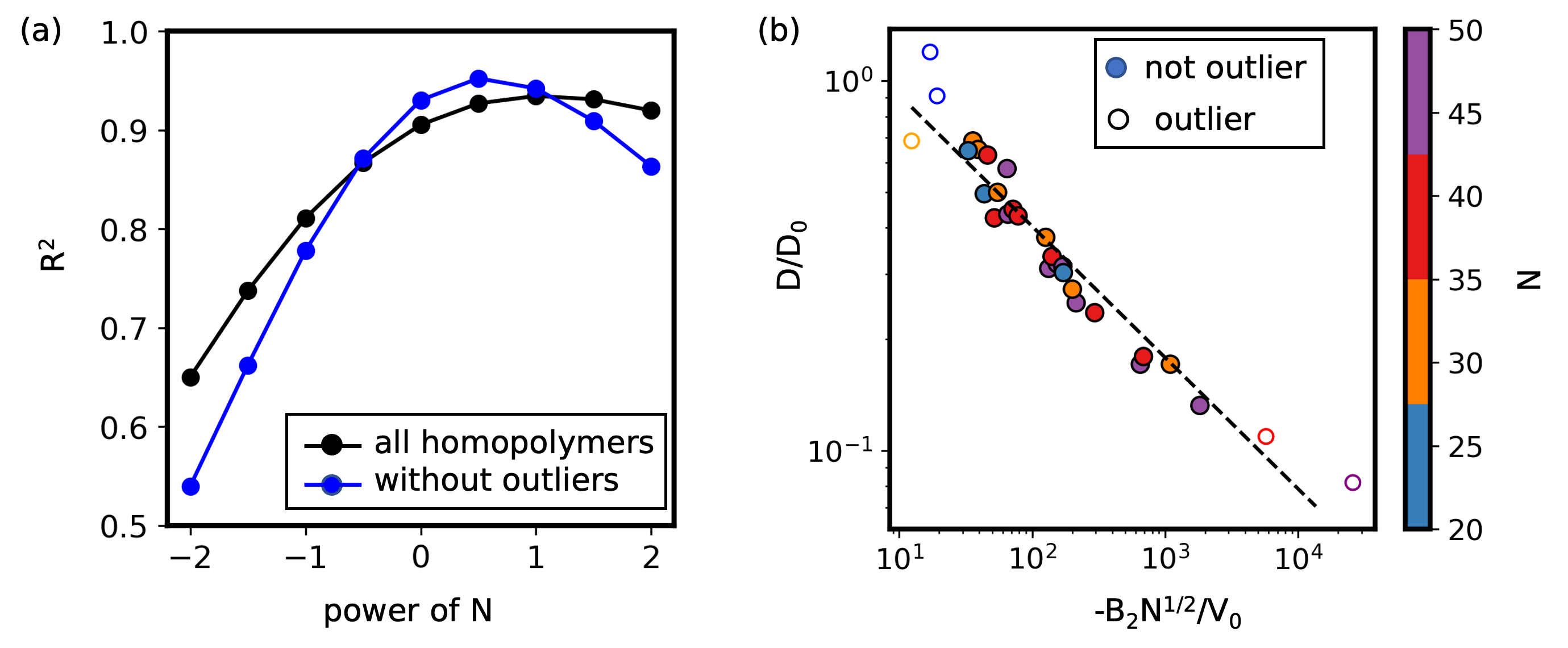}
    \caption{\textbf{Examination of thermodynamics-dynamics tradeoff for homomeric polypeptides.} \textbf{(a)} Coefficient of determination for correlation between dynamics via diffusion coefficient ($ \log [D/D_0]$) versus a measure of thermodynamic attraction via $\log[-B_2N^{\nu}/V_0]$ where $B_2$ is the second virial coefficient, $N$ is the sequence length, and $\nu$ is a power-law scaling exponent. The highest correlation is for $\nu = 1/2$ or $\nu \approx 1$. \textbf{(b)} Illustration of data collapse with $\nu = 1/2$. In both panels, data points are considered ``outliers'' if $-B_2 > 10^{6.5}$ or $-B_2 < 10^{4.5}$.}
    \label{fig:power}
\end{figure*}

\begin{figure*}
    \centering
    \includegraphics[width=0.8\linewidth]{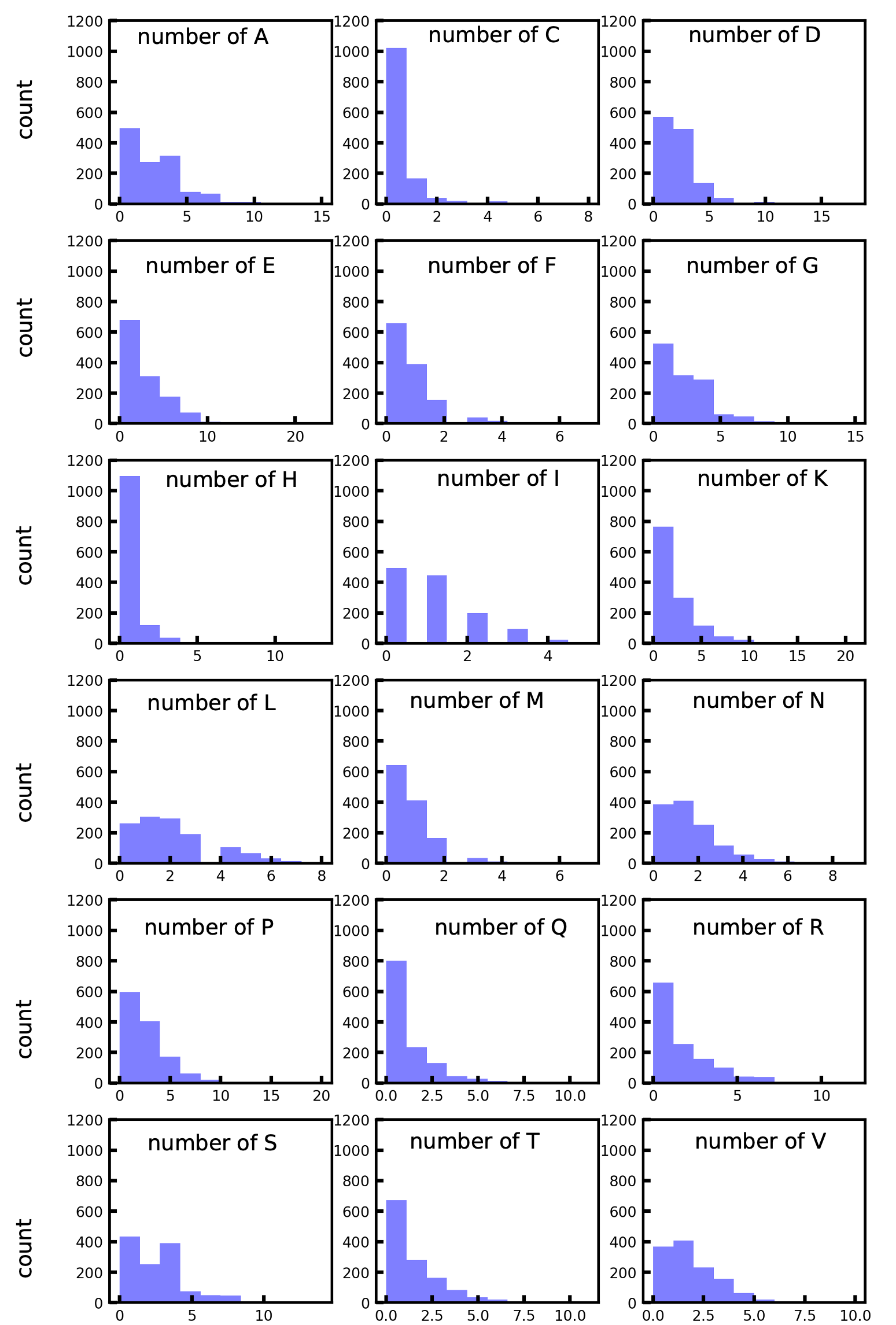}
    \caption{\textbf{Distribution of values for features among sequences selected from the DisProt database.} 
      Panels show the number of each amino-acid type (A, C, D, E, F, G, H, I, K, L, M, N, P, Q, R, S, T, V, W, or Y) and the ten sequence characteristics described in SI~\secref{sec:features}. Sequences possess lengths of $20 \le N \le 50$.}
    \label{fig:s4}
\end{figure*}

\renewcommand{\thefigure}{S\arabic{figure} (Continued)}
\addtocounter{figure}{-1}
\begin{figure*}
    \centering
    \includegraphics[width=0.8\linewidth]{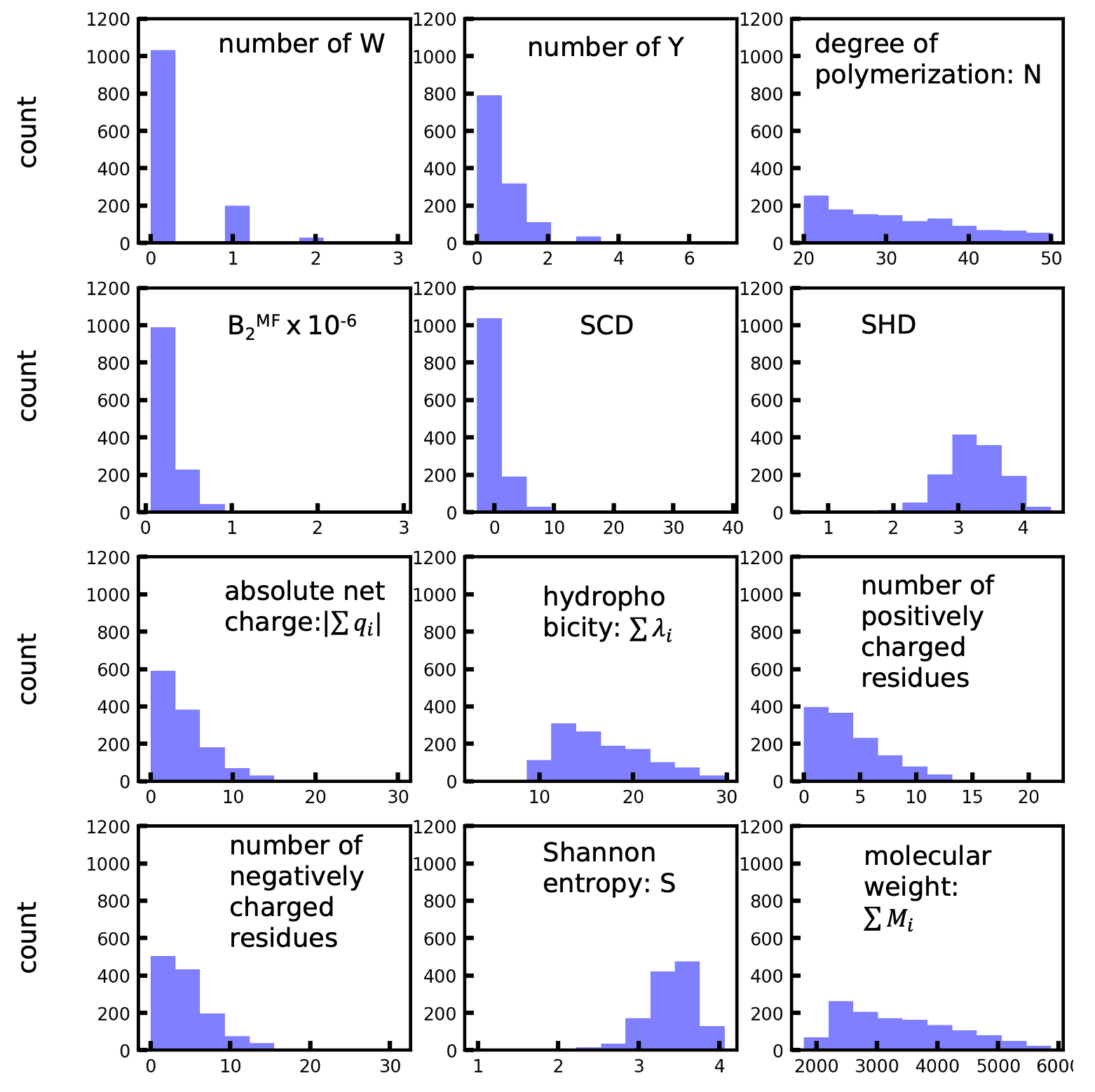}
     \caption{\textbf{Distribution of values for features among sequences selected from the DisProt database.} 
      Panels show the number of each amino-acid type (A, C, D, E, F, G, H, I, K, L, M, N, P, Q, R, S, T, V, W, or Y) and the ten sequence characteristics described in SI~\secref{sec:features}. Sequences possess lengths of $20 \le N \le 50$.}
\end{figure*}

\renewcommand{\thefigure}{S\arabic{figure}}
\begin{figure*}
\centering
\includegraphics[width=0.5\linewidth]{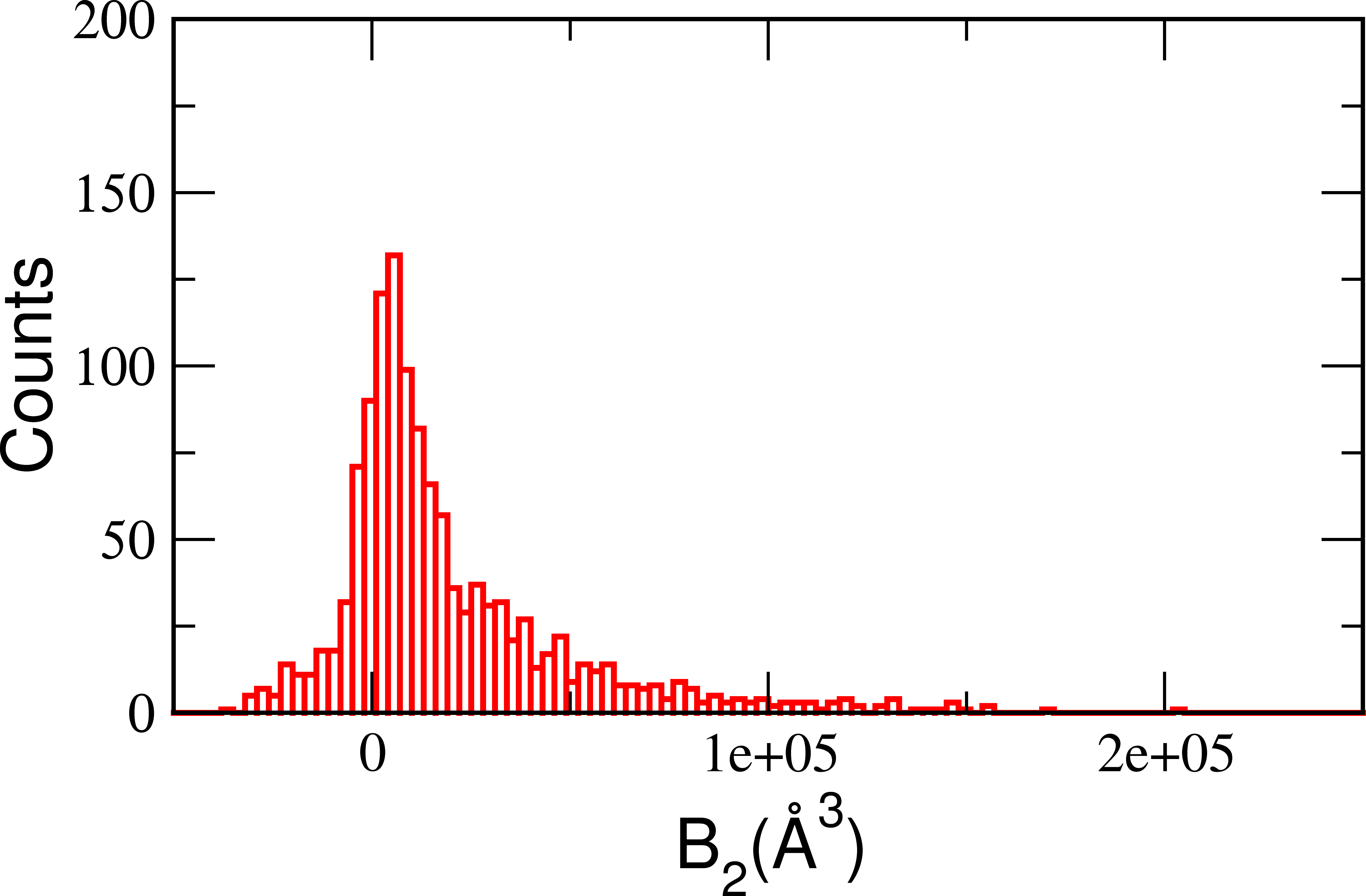}
    \caption{\textbf{Distribution of calculated second virial coefficients among sequences selected from the DisProt database.} 
      Sequences possess lengths of $20 \le N \le 50$. The majority of sequences exhibit $B_2>0$.}
    \label{fig:B2_disprot}
\end{figure*}

\begin{figure*}
    \centering
    \includegraphics[width=\linewidth]{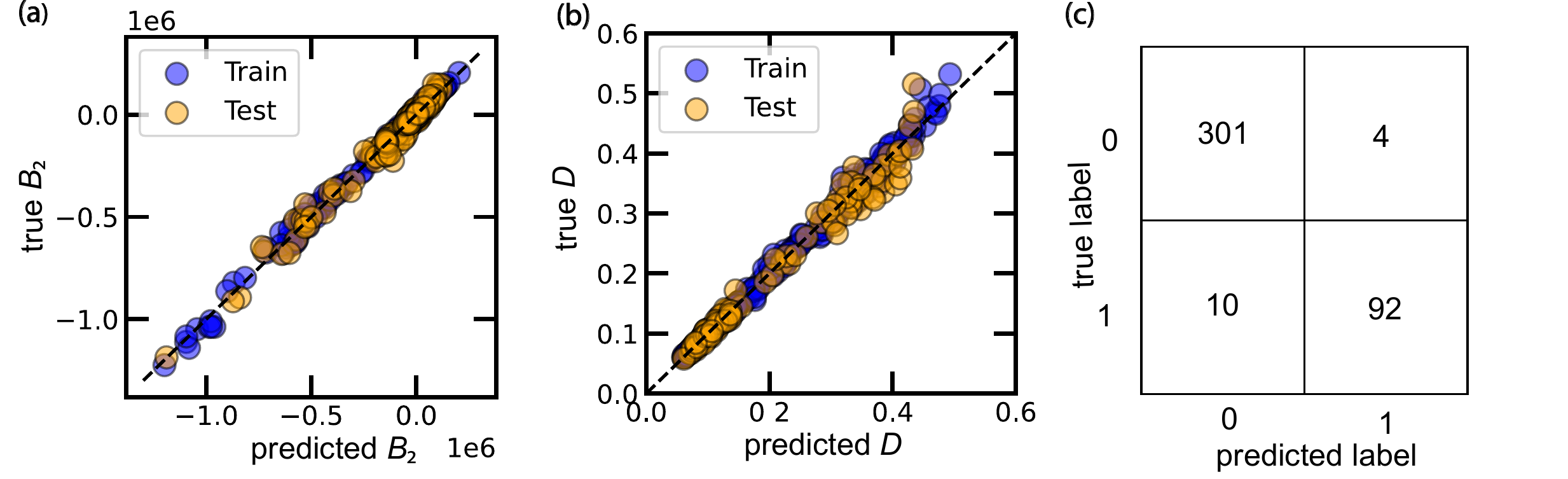}
    \caption{\textbf{Illustration of the quality of machine-learning models.} \textbf{a} Parity plot for the Gaussian process regression model that predicts the second-virial coefficient, $B_2$. 
    \textbf{(b)} Parity plot for the Gaussian process regression model that predicts the condensed-phase self-diffusion coefficient, $D$. 
    \textbf{(c)} Confusion matrix (test set) for classification of phase-separation behavior. A value of `0' indicates no phase separation and `1' indicates phase separation. The squares on the diagonal report the number of true classifications while the off-diagonal squares report the number of misclassifications. In all panels, the train-test split is 80\% for training and 20\% for testing with random selection. In \textbf{(a)} and \textbf{(c)}, all 2034 heteromeric sequences are included. In \textbf{(b)}, only the subset of the 2034 heteromeric sequences for which condensed-phase densities could be extracted were included; this includes 508 sequences. }
    \label{fig:s6}
\end{figure*}

\begin{figure}
    \centering
\includegraphics[width=0.6\linewidth]{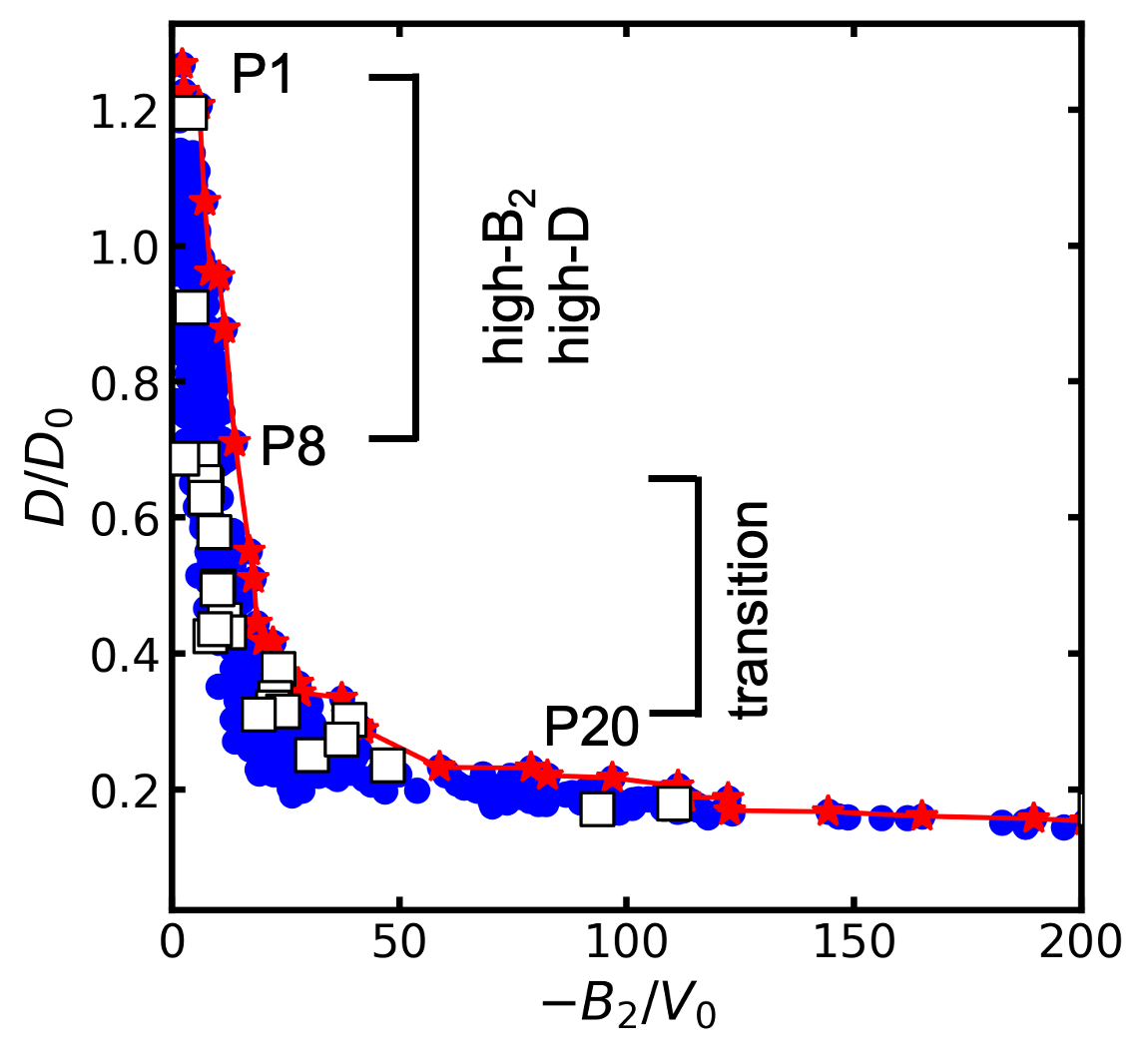}
    \caption{\textbf{The relationship between the dimensionless diffusivity, $D/D_0$, and the dimensionless second virial coefficient, $-B_2/V_0$, of all designed IDP sequences that undergo phase separation.} The plot features the same data as Fig. 2c of the main text but on linear scale and with all sequences indicated by markers. Red stars, blue circles, and white squares correspond to Pareto-optimal sequences, non-Pareto-optimal sequences, and homomeric polypeptides, respectively. The Pareto-optimal sequences P1, P8, and P20 are identified to outline the rough division of the high-$B_2$/high-D regime, transition regime, and low-$B_2$/low-$D$ regime discussed in the main text.}
    \label{fig:linear_B2_D}
\end{figure}

\begin{figure}
    \centering
     \includegraphics[width=0.4\linewidth]{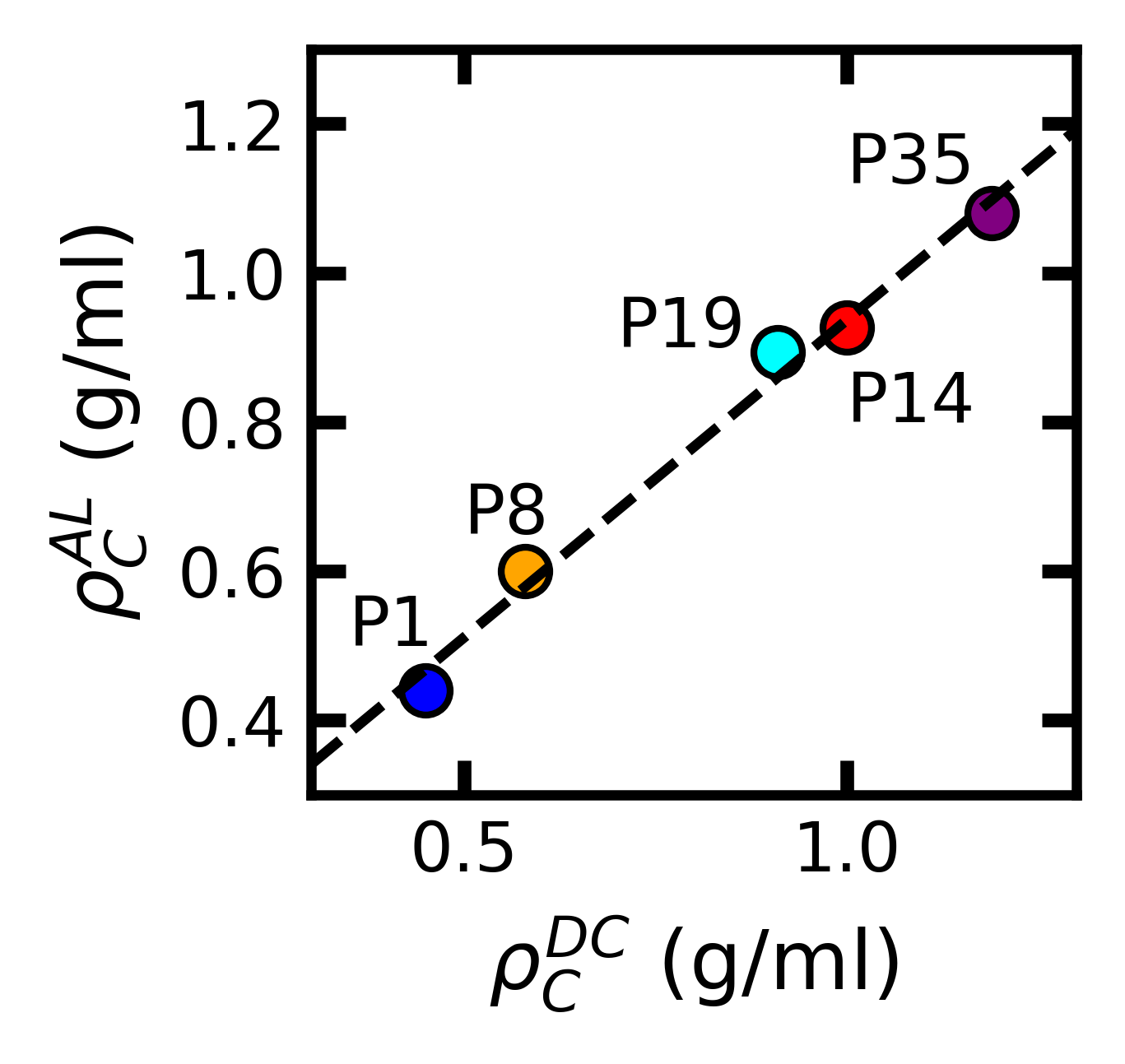}
     \caption{\textbf{Correlation of condensed-phase densities estimated  within the active-learning (AL) framework or via direct-coexistence (DC) simulations.} Within the AL framework, the equation-of-state (EOS) method is used to extract estimates for condensate density that can be used for evaluation of dynamical properties. The DC simulations correspond to those shown in Figure 3a of the main text. The dashed line shows a linear fit to the data with a coefficient of determination of $\text{R}^2 = 0.99$ and a slope of $0.85$, suggesting high correlation with consistent underestimates provided by the EOS method.}
    \label{fig:rho_al_dc}
\end{figure}

\begin{figure*}
    \centering
    \includegraphics[width=0.8\linewidth]{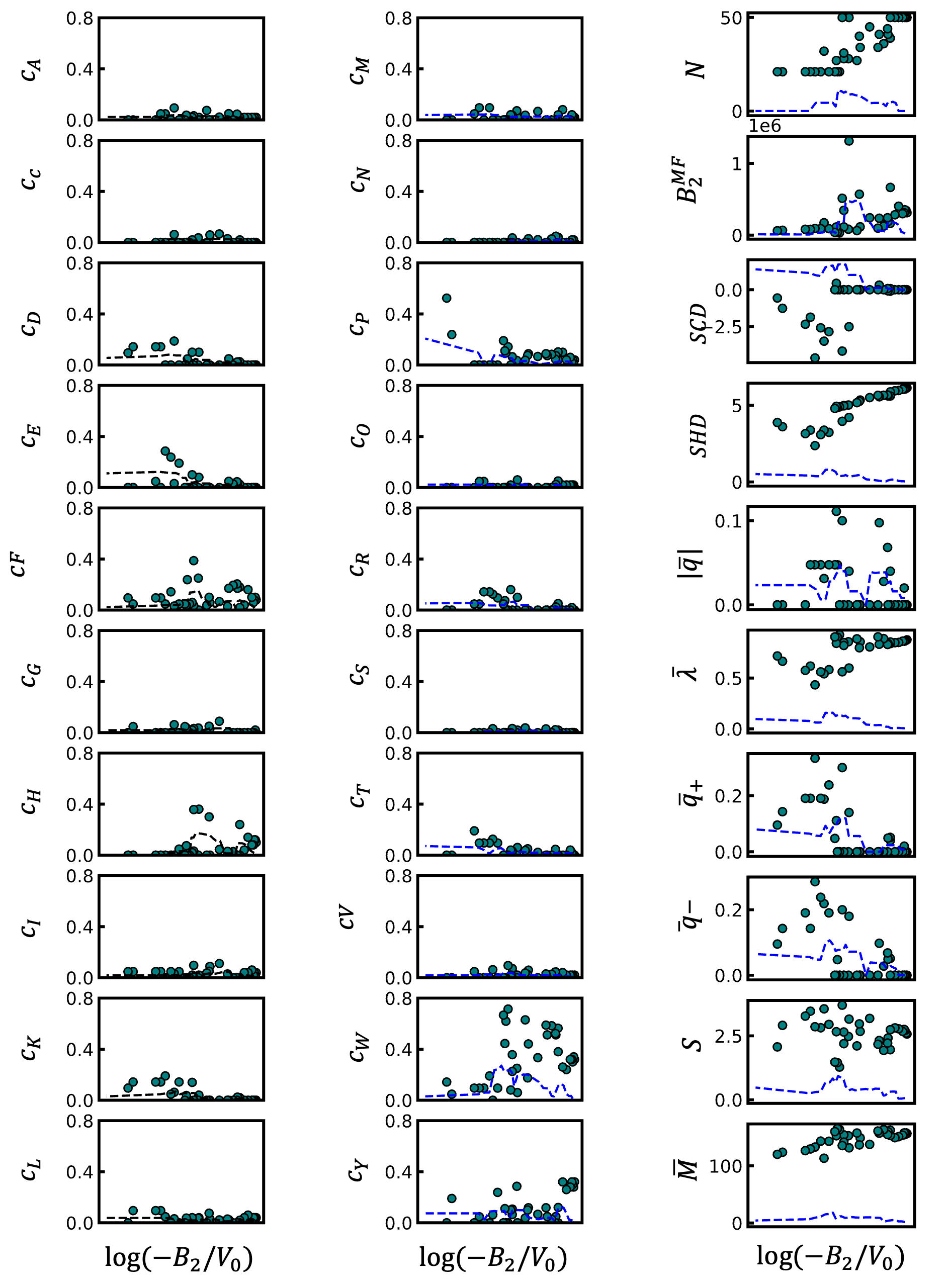}
    \caption{\textbf{Variation of feature values across the Pareto front.} Each marker indicates the value of a specific feature for a Pareto-optimal sequence. The data are presented such that both the diffusion coefficient and $-B_2$ are decreased moving left-to-right. The dashed lines indicate a rolling standard deviation of the feature $x$ (i.e., $\sqrt{\frac{1}{m}\sum_{i=1}^{m}(x_i-\bar{x})^2}$, where $\vec{x}$ is the feature vector, the bin size $m = 5$, and $\Bar{x}\equiv\frac{1}{m}\sum_{i=1}^{m} x_i$) as measure of the extent of variability of a feature in particular regions of the Pareto front.}
    \label{fig:optimal-features}
\end{figure*}

\begin{figure*}
    \centering
    \includegraphics[width=0.9\linewidth]{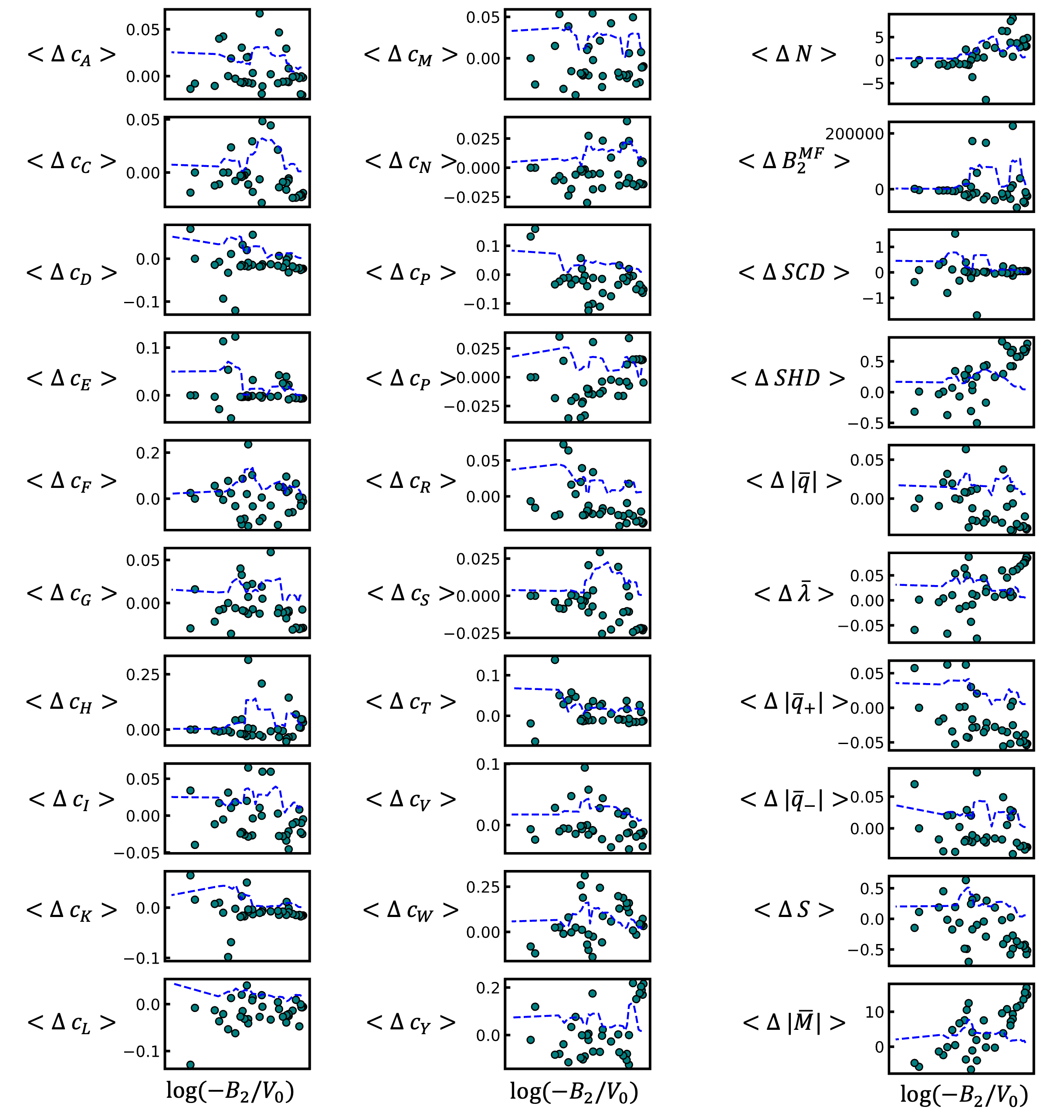}
    \caption{\textbf{Average differences in feature values between Pareto-optimal sequences and their counterfactuals.} Average feature differences determined via Eq.~(2) of the main text. The dashed lines indicate a rolling standard deviation as a measure of the extent of variability of the feature differences in particular regions of the Pareto front (see SI~\figref{fig:optimal-features}).}
    \label{fig:counterfactuals}
\end{figure*}

\begin{figure*}
    \centering
    \includegraphics[width=0.75\linewidth]{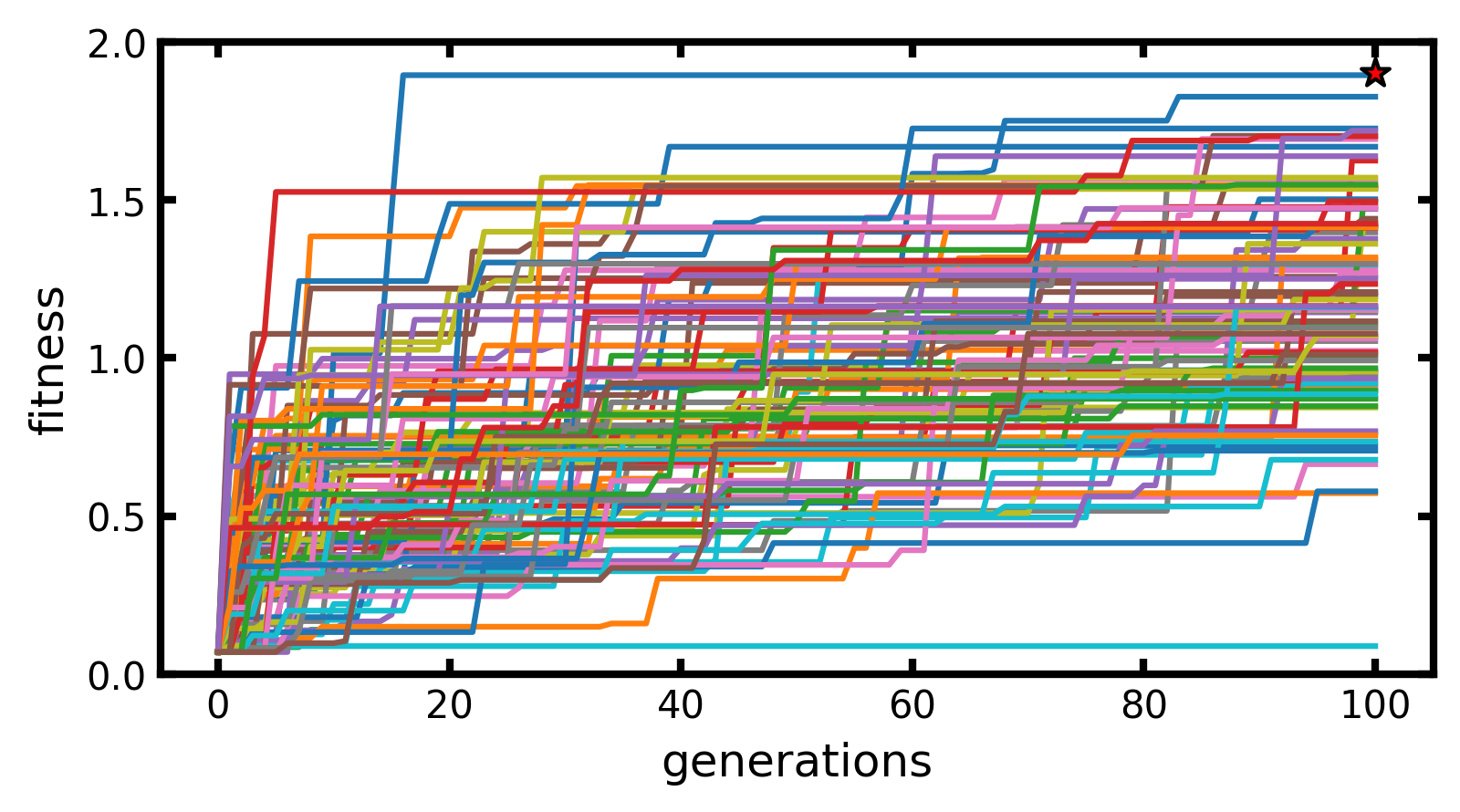}
    \caption{\textbf{Illustration of convergence toward an optimal sequence using parallel execution of the genetic algorithm.} Each solid line indicates the current highest fitness of a sequence produced over the course of 100 generations for a single execution of the genetic algorithm (\emph{Steps 1- 6} in \secref{sec:genetic-algorithm}). There are 96 lines corresponding to 96 independent executions with the same set of possible parent sequences. The different executions lead to a range of sequences with different fitness scores. The single best sequence (red star) is proposed for characterization by molecular dynamics simulation. A given active learning iteration produces 96 sequences in this manner.}
    \label{fig:convergence}
\end{figure*}

\begin{figure*}
    \centering
    \includegraphics[width=0.75\textwidth]{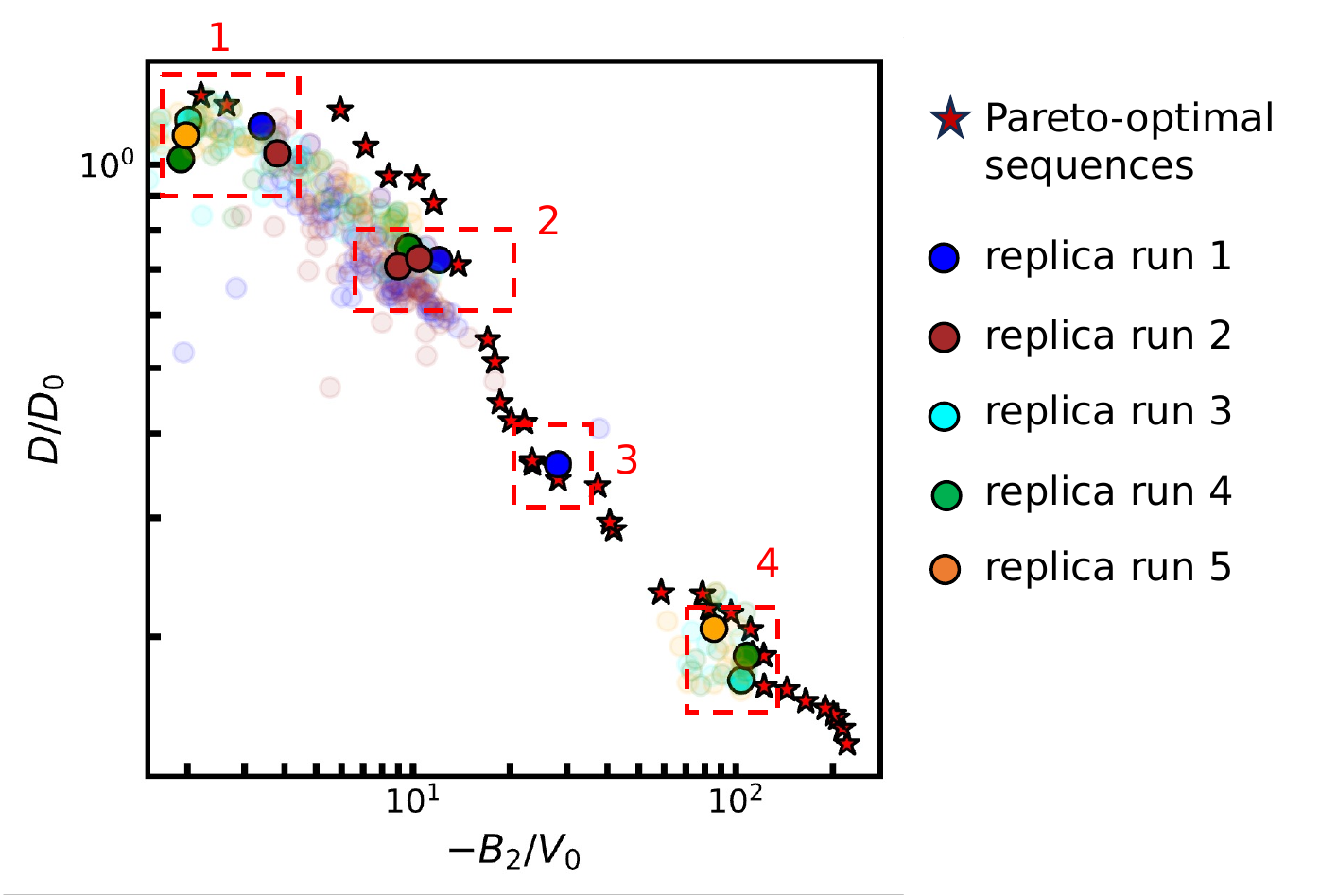}
    \caption{\textbf{Comparison of optimized sequences produced during replicate runs of the genetic algorithm.} The machine-learning predicted values of $B_2$ and $D$ for sequences generated by five replicate runs of the genetic algorithm (GA) in ``exploitation'' mode (see Figure 2c in the main text). The final Pareto front (red stars; cross referenced in Figure 2b in the main text) is shown for comparison. We compare representative sequences (opaque solid circles), each generated by a different GA replicate, from four distinct regions, 1--4, along the Pareto front to understand the stochasticity of the genetic algorithm. These replicate sequences are listed in Tables~\ref{tab:GA_b1}--\ref{tab:GA_b4}. Sequences not listed in the tables are shown as partially transparent circles for clarity.}
    \label{fig:ga_replica}
\end{figure*}

\FloatBarrier
\newpage

\begin{table}
  \centering
  \caption{Pareto-optimal sequences}
  \scriptsize
\begin{tabular}{|p{0.75cm}|p{12cm}|}
    \hline
P35 & YWWWWYWPYFHYWWWYWFMWWHIFYWLWVHYYLYWWYFYYYHPYIWYHYA \\
\hline
P34 & YWWWWYWPYFHYWWWYWFMWWHIFYWHWWHYYLYHWYFYYWQNYGWYHLF \\
\hline
P33 & YWWWWYWPYFHYWWWYWFMWWHIFYWHWVHYYLYHWYFYYYQPYIWYHLA \\
\hline
P32 & YWWWWYWPYFHYWWWYWFTWWHIYYWHWVYYYLYWWYFYYYQMNMWYHLR \\
\hline
P31 & YWWWWYWPYFFYWWWYWFPWWTIFYWHWVHYTYYHWYFWYYQPYIIYHWA \\
\hline
P30 & WFWWYFLPYFHYFWWFWFIWWHHFYWHWVHYYLYHWYFYYYQPYIWYHLA \\
\hline
P29 & YWWWWYWPYFPYWWWYWPMWYMIFYWHYVPYYLYYWYMYVYQPYHWIMLA \\
\hline
P28 & MHWYWHWNHWWNWYHHWWYWKYSHPWWFLWAPRWHWWHHHMYYWWHWWCH \\
\hline
P27 & RQFWFPEDRFQWFFWWWWWWWFWPWWWWWWWFWPWWWWP \\
\hline
P26 & DKRQFFEWWWTWNFWYFFYNFWWWWWWWFWPWWWWPWWWWP \\
\hline
P25 & AEEDTVWAFFWVFWWWSWFWFWPWTFWFFWWWWWWWFWPWWWWW \\
\hline
P24 & VHFWFYWWWNWWWPWPEWWFWFWFFWWWWWWWFWWP \\
\hline
P23 & DFLEFDFYWEFWWHWYAWFAPWWWWFWWWWWWWFWPWWWWP \\
\hline
P22 & PWWFWWWLWWWWWSWPWWWHWYCWYWPVWIYWWY \\
\hline
P21 & GFWPYGPIWWWWWCIWIIWYLFGFWMWHWANIMHCGWWWWYPMCT \\
\hline
P20 & WIWNWGCLVIFWWPWPWWWFWYCWYWPVWIYWWY \\
\hline
P19 & GPHWVHWPHHPYHHWWWHHLTLFYWHIYFHGWFFIHTHYL \\
\hline
P18 & IPWWWWFWWWWWATWWWWYMWWSWWVA \\
\hline
P17 & HEWDDEDWTQHHHHHAHEHHHHRTIEWQTDHDVLRHHHKQHMSRKLHRHR \\
\hline
P16 & WFWWYYLWYYMWWYMFGFFYYPFFWYFC \\
\hline
P15 & MHWYWHWNHWWHWYHHWWHFVWVHHYWH \\
\hline
P14 & FWWFWPWIGFFFWIFYSFFFTWIFPHWFFYA \\
\hline
P13 & YEDDDFERDEDKWYGTFEFLWKHMRQYYKCREKARIWKYVRRWKLKMRRT \\
\hline
P12 & PWWFWWWPWWWWWVVWPWWWW \\
\hline
P11 & DFFWFFWWWWWWWFWPWWWWP \\
\hline
P10 & RAYWWHGWPWWWPLPWYWWFHWRWWYK \\
\hline
P9 & PPWPGPWWWWRWWWFWWWWWW \\
\hline
P8 & EEEEWIHYTTYWYYFYKKRRK \\
\hline
P7 & TKRGRTARKTGALCRTYFACYSDDDVDVDIDE \\
\hline
P6 & RRRKMTFWWWTFWIEEFMEEE \\
\hline
P5 & RRRKKKFKWQTLWAEETEEEE \\
\hline
P4 & DDDMQAFLTFTWWYLIMRKKK \\
\hline
P3 & DDEMDTTLTFTWWFLIVRKKK \\
\hline
P2 & KKKLPYLPIGYFWYDPPDYPD \\
\hline
P1 & KKPPWPWPFPIPDPWFPPPPD \\
\hline
    \end{tabular}
    \label{tab:pareto_seq}
\end{table}

\begin{table}
  \centering
  \caption{Representative sequences in the region 1 of \figref{fig:ga_replica} generated by five GA replicates. The Pareto-optimal sequence P1 study is shown for comparison.}
  \scriptsize
    \begin{tabular}{|p{1.5cm}|p{12cm}|}
    \hline
        \textbf{Replicate} & \textbf{Sequence} \\
    \hline
        P1 & KKPPWPWPFPIPDPWFPPPPD\\
    \hline
        1 & IGLRAQTWPPMWQQWILPQPD\\
    \hline
        2 & DWWEDLQYPQPFRKQMWRWEK\\
    \hline
        3 & KKPTWPWPGPIPFPWFPSPPF\\
    \hline
        4 & RRRCKKMDYTHMPDLITDTED\\
    \hline
        5 & YDNTFRQCTPGMWFIVCLWMM\\
    \hline
    \end{tabular}
    \label{tab:GA_b1}
\end{table}
\begin{table}
    \centering
    \caption{Representative sequences in the region 2 of \figref{fig:ga_replica} generated by five GA replicates. The Pareto-optimal sequence P8 is shown for comparison.}
    \scriptsize
    \begin{tabular}{|p{1.5cm}|p{12cm}|}
    \hline
        \textbf{Replicate} & \textbf{Sequence} \\
    \hline
        P8 & EEEEWIHYTTYWYYFYKKRRK\\
    \hline
        1 & TWRPRTARKTFAMPYTYWYCYSDDDVDEDNL\\
    \hline
        2 & DDEGDCGPCDDCMPFKPEFVLLTFTWWFLIVRKKK\\
    \hline
        2 & DDDDIPQLWDSRTFKILKFLIWFLIVRKKK\\
    \hline
        4 & RRRKKKLRSKKKFCWQTEWAEETEEEE\\
    \hline
    \end{tabular}
    \label{tab:GA_b2}
\end{table}
\begin{table}
    \centering
    \caption{Representative sequences in the region 3 of \figref{fig:ga_replica} generated by five GA replicates. The Pareto-optimal sequences P14 and P15 are shown for comparison.}
    \scriptsize
    \begin{tabular}{|p{1.5cm}|p{12cm}|}
    \hline
        \textbf{Replicate} & \textbf{Sequence} \\
    \hline
       P14 & FWWFWPWIGFFFWIFYSFFFTWIFPHWFFYA\\
    \hline
        P15 & MHWYWHWNHWWHWYHHWWHFVWVHHYWH\\
    \hline
        1 & DDNTWDWNHMVPEHHDDVIFHWNHWWHWYHHWWHWWHFVWVHHYWH\\
    \hline
    \end{tabular}
    \label{tab:GA_b3}
\end{table}
\begin{table}
    \centering
    \caption{Representative sequences in the region 4 of \figref{fig:ga_replica} generated by five GA replicates. The Pareto-optimal sequences P21--P24 are shown for comparison.}
    \scriptsize
    \begin{tabular}{|p{1.5cm}|p{12cm}|}
    \hline
        \textbf{Replicate} & \textbf{Sequence} \\
    \hline
         P21 &  GFWPYGPIWWWWWCIWIIWYLFGFWMWHWANIMHCGWWWWYPMCT\\
    \hline
         P22 & PWWFWWWLWWWWWSWPWWWHWYCWYWPVWIYWWY \\
    \hline
         P23 & DFLEFDFYWEFWWHWYAWFAPWWWWFWWWWWWWFWPWWWWP\\
    \hline
        P24 & VHFWFYWWWNWWWPWPEWWFWFWFFWWWWWWWFWWP\\
    \hline
        3 & QRWFPWWKESTWWESHWWWFWYCRWWFWWWLWWWWWYWPVWIYWWY\\
    \hline
        4 & WWRHWYWPPHGQYQWDLQPWHWLPWVYWWHWYWWWFWLYIYPFYWPHYW\\
    \hline
        5 & QAWRWHWIRWRIWWHWWAYHQHWWWWRWWWFWWWWWWWWW\\
    \hline
    \end{tabular}
    \label{tab:GA_b4}
\end{table}


\newgeometry{margin=1.8cm} 
\begin{landscape}

  \scriptsize{
\begin{longtable}{|p{0.6cm}|p{10.8cm}|p{0.6cm}|p{10.8cm}|}
\caption{Counterfactual sequences for Pareto-optimal sequences}\label{tab:counterfactuals}\\\hline
\endfirsthead
\caption*{Table \ref{tab:counterfactuals} continued:}\\\hline
\endhead
\endfoot
\hline\endlastfoot
$\mathcal{C}_{ 1,1 }$ & PWLYWWWPFPWPPIFPWPPPP&  $\mathcal{C}_{ 1,2 }$ & FFCWPWPLPLPPPLPPFFVWP \\
$\mathcal{C}_{ 1,3 }$ & DLDLLLPYAWPWLLWYLFLKK&  $\mathcal{C}_{ 1,4 }$ & PWPLWAWLPLPTGPPPWCPLP \\
$\mathcal{C}_{ 1,5 }$ & LLPWFTPPPWWWLWWPPWGPP&  $\mathcal{C}_{ 1,6 }$ & PPIPPPPWWDWPWPWPWKPPR \\
$\mathcal{C}_{ 1,7 }$ & KKPPGPPPGPWWWWFPCTGFWFPPPPD & & \\ 
\hline 
$\mathcal{C}_{ 2,1 }$ & DDDPFLTYYWWFYPTFGLKKK&  $\mathcal{C}_{ 2,2 }$ & DDDLYLWFYWWPLPPWPRKPK \\
$\mathcal{C}_{ 2,3 }$ & DDGDWWWIWGTWTPPKPKKWW&  $\mathcal{C}_{ 2,4 }$ & DDDIWIIWVYIIIILKIRKIT \\
$\mathcal{C}_{ 2,5 }$ & KKKVWMWLMMLDLWLLMDLDV&  $\mathcal{C}_{ 2,6 }$ & DDDGTAFWTFTWYYIWLKKKY \\
\hline 
$\mathcal{C}_{ 3,1 }$ & TWEEEEYDWYTQWMNKKKKYK&  $\mathcal{C}_{ 3,2 }$ & LDDDDWGLLCYLYFWCTRKKK \\
\hline
\end{longtable}
}

\end{landscape}
\restoregeometry